\documentclass[usegraphicx,usenatbib]{mn2e}

\newcommand{\subs}[1]{$_{\rm #1}$}
\newcommand{\sups}[1]{$^{\rm #1}$}
\newcommand{\degree}{$^\circ$}
\newcommand{\degrees}{$^\circ$ }
\newcommand{\BE}{\begin{equation}}
\newcommand{\EE}{\end{equation}}
\newcommand{\kmsn}{km\ s$^{-1}$}
\newcommand{\kmss}{km\ s$^{-1}$ }
\newcommand{\vsinis}{$\!${\em v\,}sin{\em i} }
\newcommand{\vsini}{$\!${\em v\,}sin{\em i}}
\def\ga{\mathrel{\hbox{\rlap{\hbox{\lower4pt\hbox{$\sim$}}}\hbox{$>$}}}}
\def\la{\mathrel{\hbox{\rlap{\hbox{\lower4pt\hbox{$\sim$}}}\hbox{$<$}}}}

\title[The magnetic field of HD 141943]{Magnetic fields and differential rotation on the pre-main sequence I: The early-G star HD 141943 - brightness and magnetic topologies}
\author[S. C. Marsden et al.]
  {S.~C.~Marsden,$^{1,2}$\thanks{Email: Stephen.Marsden@jcu.edu.au (SCM); mmj@st-and.ac.uk (MMJ); jramirez@astroscu.unam.mx (JCRV); evelyne.alecian@obs.ujf-grenoble.fr (EA); brownca@usq.edu.au (CJB); carterb@usq.edu.au (BDC); donati@obs-mip.fr (J-FD); n.j.dunstone@googlemail.com (ND); rhodes.hart@usq.edu.au (RH); Meir.Semel@obspm.fr (MS); waite@usq.edu.au (IAW)} M.~M.~Jardine,$^3$\footnotemark[1] J.~C.~Ram\'{i}rez V\'{e}lez,$^{4,5}$\footnotemark[1]  E.~Alecian,$^{4,6}$\footnotemark[1] \and C.~J.~Brown,$^7$\footnotemark[1] B.~D.~Carter,$^7$\footnotemark[1] J.-F.~Donati,$^8$\footnotemark[1] N.~Dunstone,$^3$\footnotemark[1] R.~Hart,$^7$\footnotemark[1] \and M.~Semel$^4$\footnotemark[1]  and I.~A.~Waite$^7$\footnotemark[1]\\
   $^1$Centre for Astronomy, School of Engineering and Physical Sciences, James Cook University, Townsville, 4811, Australia\\
   $^2$Australian Astronomical Observatory, PO Box 296, Epping NSW 1710, Australia\\
   $^3$SUPA, School of Physics and Astronomy, University of St. Andrews, St. Andrews, KY 16 9SS, Scotland\\ 
   $^4$LESIA, Observatoire de Paris-Meudon, F-92195 Meudon Cedex, France\\
   $^5$Instituto de Astronomia, Universidad Nacional Aut\'{o}noma de M\'{e}xico, 04510, Coyoacon, M\'{e}xico D.F.\\
   $^6$LAOG, Laboratoire d'Astrophysique de Grenoble, Universit\'{e} Joseph Fourier, BP 53 38041, Grenoble Cedex 09, France\\
   $^7$Faculty of Sciences, University of Southern Queensland, Toowoomba, 4350, Australia\\
   $^8$LATT--UMR 5572, CNRS \& Univ.\ de Toulouse, 14 Av.\ E.~Belin, F--31400 Toulouse, France}
\date{Accepted version}

\pagerange{\pageref{firstpage}--\pageref{lastpage}} \pubyear{2010}

\begin{document}

\label{firstpage}

\maketitle

\begin{abstract}
Spectroscopic and spectropolarimetric observations of the pre-main sequence early-G star HD 141943 were obtained at four observing epochs (in 2006, 2007, 2009 and 2010). The observations were undertaken at the 3.9-m Anglo-Australian Telescope using the UCLES echelle spectrograph and the SEMPOL spectropolarimeter visitor instrument. Brightness and surface magnetic field topologies were reconstructed for the star using the technique of least-squares deconvolution to increase the signal-to-noise of the data. 

The reconstructed brightness maps show that HD 141943 had a weak polar spot and a significant amount of low latitude features, with little change in the latitude distribution of the spots over the 4 years of observations. The surface magnetic field was reconstructed at three of the epochs from a high order ($l$ $\le$ 30) spherical harmonic expansion of the spectropolarimetric observations. The reconstructed magnetic topologies show that in 2007 and 2010 the surface magnetic field was reasonably balanced between poloidal and toroidal components. However we find tentative evidence of a change in the poloidal/toroidal ratio in 2009 with the poloidal component becoming more dominant. At all epochs the radial magnetic field is predominantly non-axisymmetric while the azimuthal field is predominantly axisymmetric with a ring of positive azimuthal field around the pole similar to that seen on other active stars.
\end{abstract}


\section{Introduction} \label{Sec_int}

The generation of magnetic fields is arguably one of the most important process operating inside a star affecting everything from the angular momentum evolution of the star through to the habitability of any planets around the star. In the solar case the magnetic dynamo is believed to operate in an interface layer between the differentially rotating convective zone and the radiative zone which rotates as a solid-body \citep{ParkerEN:1993}. However, for young, rapidly-rotating solar-type stars evidence is growing that such stars may in fact have a different dynamo mechanism. It has been suggested \citep{DonatiJF:2003} that such stars may in fact house distributed dynamos, i.e.\ dynamos which operate across the entire convective zone, rather than being restricted to the interface-layer as in the solar case. The strongest evidence for this are large regions of near-surface azimuthal magnetic field seen on such stars \citep[i.e.][]{DonatiJF:2003, MarsdenSC:2006a}. Such regions are believed to be the near-surface toroidal components of the large-scale dynamo field. In a solar-like interface-layer dynamo such fields should not be seen near the stellar surface and thus it is postulated that a distributed dynamo is operating in young, rapidly-rotating solar-type stars.

Most current dynamo models are based on our understanding of the Sun and are tailored to reproduce solar observations \citep[see][]{ParkerEN:1993, CharbonneauP:2005}. Such models usually involve an interface layer dynamo which, as discussed, may not apply to young solar-type stars, but there are some different dynamo models that have been examined. For example, the dynamo operating in fully convective stars \citep[i.e.][]{BrowningMK:2008}, or near-surface dynamos \citep[i.e.][]{BrandenburgA:2005} or rapidly-rotating solar models with no interface layer \citep[i.e.][]{BrownBP:2010}. Still the operation of stellar magnetic dynamos  is not well understood.

One of the most direct ways of observing the stellar dynamo is through the observation of the global magnetic field on the surface of a star. Zeeman Doppler imaging \citep[ZDI,][]{SemelM:1989, DonatiJF:1997b}  has been used for a number of years now to observe the magnetic field configurations of solar-type stars \citep[i.e.][]{DonatiJF:1992, DonatiJF:1999a, DonatiJF:1997a, DonatiJF:1999b, PetitP:2004a, PetitP:2004b, PetitP:2008, MarsdenSC:2006a, JeffersSV:2008, DunstoneNJ:2008} and there is now a growing body of observations of the surface topologies of other stars that any stellar dynamo models needs to be able to reproduce. 

\citet{DonatiJF:2009} have summarised many of the observations of the magnetic fields of non-degenerate stars and for young solar-type stars (i.e.\ those with Rossby numbers $\la$ 1 and more massive than $\sim$0.5 M\subs{\odot}) they find that these stars produce substantial toroidal fields and have mostly non-axisymmetric poloidal fields. For more mature stars, \citet{PetitP:2008} have shown that for solar-type stars with rotation periods faster than $\sim$12 days the global magnetic field of the star appears to be dominated by toroidal field with poloidal field being by far the most dominant field configuration for stars with slower rotation rates. This could indicate a change in the dynamo mechanism for solar-type stars rotating faster than $\sim$12 days. Such observations have yet to be modelled in stellar dynamo theory.

Doppler images of the surface spot topology of young solar-type stars show that they almost universally have large polar spot features \citep[i.e.][]{BarnesJR:2000, BarnesJR:2001a, BarnesJR:2001b, DonatiJF:2003, MarsdenSC:2005b, MarsdenSC:2006a}. This is in contrast to the predictions of current dynamo theory \citep[i.e.][]{SchusslerM:1996, GranzerT:2000, GranzerT:2004} and it has been suggested that a strong meridional flow may be responsible for transporting these spots from their emergence latitudes to the polar regions. This may also explain the mixed polarity of magnetic fields seen on the polar regions of many stars \citep[see][]{MackayDH:2004}.

Although great progress has been made, our knowledge of stellar magnetic dynamos is still in its infancy. How the operation of the stellar magnetic dynamo depends on basic stellar parameters such as age, mass and rotation rate is still unknown.

One of the most obvious effects of the solar dynamo is the reversal of the Sun's global magnetic topology every $\sim$11 years, but so far only two other solar-type stars have shown evidence of global magnetic polarity reversals, HD 190771 \citep{PetitP:2009} and Tau Boo \citep{DonatiJF:2008, FaresR:2009}. No strong evidence for a global polarity reversal has yet been seen on a young solar-type star, even though some, for example AB Dor \citep{DonatiJF:1997a, DonatiJF:1999a, DonatiJF:2003} have been observed for a number of years. Thus it is still unknown if young solar-type stars undergo regular magnetic cycles like the Sun or have chaotic cycles, as indicated by the Calcium HK emission of young stars \citep{BaliunasSL:1995}. 

For young solar-type stars the available spectropolarimetric observations are mainly of K-stars \citep[i.e.][]{DonatiJF:2003}. At present there is only one young early-G star for which spectropolarimetric observations have been published in a refereed journal, HD 171488 \citep{MarsdenSC:2006a, JeffersSV:2008, JeffersSV:2010}, with preliminary results for the late-F star HR 1817 \citep{MengelM:2006, MarsdenSC:2006b, MarsdenSC:2010a} being published as conference proceedings.

As part of a study into the magnetic topology and cycles of young late-F/early-G stars this paper along with that of \citet[][Paper II]{MarsdenSC:2010b} and \citet{WaiteIA:2010}, presents spectropolarimetric observations of two early-G pre-main sequence (PMS) stars. This paper and Paper II deals with the observations of the young early-G star HD 141943, while the \citet{WaiteIA:2010} paper deals with observations of the similarly aged but more massive HD 106506.

HD 141943 was first identified by \citet{WaiteIA:2005} as a potential target for spectropolarimetric observations. It is active \citep[Log(L\subs{X}) = 30.7 erg s\sups{-1},][]{CutispotoG:2002} and bright \citep[V = 7.9,][]{CutispotoG:2002}. It also has a very strong Lithium line \citep[A\subs{Li} = 3.3,][]{CutispotoG:2003} indicating its youth \citep[$\sim$15 Myrs,][]{CutispotoG:2003} and it is what we class as a moderately rapid rotator with a \vsinis $\sim$38 \kmss \citep{CutispotoG:2003} and a period of 2.20 $\pm$ 0.03 days \citep{CutispotoG:1999}. According to \citet{HillenbrandLA:2008} HD 141943 is also possibly host to a $\sim$85 K debri disk.

This paper (Paper I) describes the evolution in both the brightness (reconstructed at 4 epochs) and magnetic (reconstructed at 3 epochs) topologies of HD 141943 taken at the Anglo-Australian telescope (AAT). A further paper on HD 141943 (Paper II) discusses the differential rotation, H$\alpha$ emission and coronal magnetic field maps reconstructed from these observations.

\section{Observations and data reduction} \label{Sec_obs}

HD141943 was observed at 4 epochs on the 3.9-m AAT, in May 2006, March/April 2007, April 2009 and March/April 2010. The May 2006 observations were standard spectroscopic observations taken over a 6 night period using the University College London Echelle Spectrograph (UCLES). The 2007, 2009 and 2010 observations were spectropolarimetric, observing both left- and right-hand circularly polarised light using the SEMPOL (or SEMELPOL) spectropolarimeter \citep*{SemelM:1993, DonatiJF:2003} visitor instrument in conjunction with UCLES. The observations in 2007, 2009 and 2010 were taken over 11, 7 and 12 nights respectively. A log of the observations is given in Table~\ref{Tab_log}. The rotational phase of the observations was determined from the following ephemeris:
\begin{equation}
{\rm HJD} = 2454195.154 + 2.182\phi, \label{Eqn_eph}
\end{equation}
where HJD is the Heliocentric Julian date of the observation and $\phi$ is the rotational phase.  

\begin{table*}
\caption{Logs of spectroscopic and spectropolarimetric AAT observations of HD 141943 for May 2006, March/April 2007, April 2009 and March/April 2010. The first two columns give the UT dates and times of the mid point of each observation, while the third column gives the exposure time. The fourth column gives the rotational phase of the observations calculated from equation~\ref{Eqn_eph}. For the May 2006 observations the rotational phases have had 153.0 added to them, the March/April 2007 rotational phases have had 3.0 added to them, the April 2009 rotational phases have had 336.0 subtracted from them, while for the March/April 2010 observations the rotational phases have had 497.0 subtracted from them. This is so that all phases are shown as positive. The fifth column gives the signal-to-noise ratio of the resultant Stoves V LSD profile (only for the 2007, 2009 and 2010 spectropolarimetric data).}
\label{Tab_log}
\centering
\begin{tabular}{ccccc}
\hline\hline
UT date      & UT time       & Exp.                   & Rot. phase  &                             \\
                    &                      &   (sec.)                & (+ 153.0)     &                             \\
\hline
06 May 06 & 18:44:03 to & 400, 300,           & 0.448 to      &                             \\
                    & 19:46:17     & 5$\times$600    & 0.468          &                             \\
07 May 06 & 17:48:15 to & 9$\times$600   & 0.889 to      &                             \\
                    & 19:17:00     &                             & 0.917          &                              \\
08 May 06 & 17:16:24 to & 13$\times$600 &  1.337 to    &                              \\
                    & 19:27:28     &                             &  1.379         &                              \\
09 May 06 & 17:26:23 to & 13$\times$600 &  1.798 to    &                              \\
                    & 19:38:43     &                              &  1.840        &                              \\
10 May 06 & 16:53:01 to & 15$\times$600 &  2.246 to    &                               \\
                    & 19:30:00     &                             &  2.296         &                              \\
11 May 06 & 17:39:12 to & 11$\times$600 &  2.719 to    &                               \\
                    & 19:27:05     &                             &  2.753         &                              \\
\hline\hline 
UT date     & UT time       & Exp.                     & Rot. phase & S/N\subs{LSD} \\
                   &                      & (sec.)                   & (+ 3.0)         &                              \\
\hline
30 Mar 07 & 12:30:43     & 4$\times$600    & 0.649          & 4948                    \\
31 Mar 07 & 12:20:25     & 4$\times$600    & 1.104          & 4507                    \\
01 Apr 07 & 12:26:09     & 4$\times$600    & 1.564           & 4656                    \\
                   & 18:40:53     & 4$\times$600    & 1.684           & 4544                   \\
02 Apr 07 & 12:23:33     & 4$\times$600    & 2.022           & 5959                   \\
                   & 18:14:40     & 4$\times$600    & 2.134           & 7715                   \\
03 Apr 07 & 12:26:18      & 4$\times$600    & 2.481          & 4606                   \\
                   & 18:56:19     & 4$\times$600    & 2.605           & 5354                   \\
04 Apr 07 & 13:07:40     & 4$\times$600    & 2.953           & 5009                   \\
                   & 18:58:10     & 4$\times$600    &  3.064          & 6980                   \\
05 Apr 07 & 12:09:39     & 4$\times$600    &  3.392          & 4873                    \\
                   & 15:42:02     & 4$\times$600    &  3.460          & 4599                   \\
                   & 18:40:21     & 4$\times$600    &  3.517          & 6617                   \\
06 Apr 07 & 12:33:12     & 4$\times$600    &  3.858          & 4915                   \\
                   & 16:18:35     & 4$\times$600    &  3.930          & 6553                   \\
                   & 18:35:54     & 4$\times$600    &  3.974          & 6700                   \\
07 Apr 07 & 12:18:45     & 4$\times$600    &  4.312          & 7035                   \\
08 Apr 07 & 12:29:38     & 4$\times$600    &  4.774          & 6963                   \\
                   & 16:11:27     & 4$\times$600    &  4.844          & 6734                   \\
                   & 18:47:16     & 4$\times$600    &  4.894          & 6847                   \\
09 Apr 07 & 12:18:25     & 4$\times$600    &  5.228          & 5992                   \\
                   & 16:11:12    & 4$\times$600    &  5.303           & 7662                   \\
                   & 18:40:21    & 4$\times$600    &  5.350          & 8100                    \\
                    &                    &                              &                      &                              \\
\hline
\end{tabular}
\begin{tabular}{ccccc}
\hline\hline
UT date    & UT time   & Exp.                 & Rot. phase & S/N\subs{LSD} \\
                   &                  & (sec.)               & (- 336.0)     &                             \\
\hline
07 Apr 09 & 13:31:28 & 4$\times$600 & 0.349         & 3827                  \\
                    & 16:16:43 & 4$\times$600 & 0.405         & 3327                  \\
                    & 18:45:33 & 4$\times$600 & 0.449         & 2251                   \\
08 Apr 09 & 13:25:37 & 4$\times$600 & 0.805         & 5216                   \\
                   & 18:51:41 & 4$\times$800 & 0.909         & 7225                   \\  
09 Apr 09 & 13:59:27 & 4$\times$800 & 1.274         & 5050                   \\
                   & 17:07:59 & 4$\times$800 & 1.334         & 5760                   \\ 
10 Apr 09 & 14:33:60 & 2$\times$800 & 1.744         & 5902                   \\ 
13 Apr 09 & 13:38:03 & 2$\times$800 &  3.101       & 4730                    \\
                   &                  &                           &                    &                              \\
                   &                  &                           &                    &                              \\
                   &                 &                           &                    &                               \\
\hline\hline
UT date     & UT time   & Exp.                  & Rot. phase & S/N\subs{LSD} \\ 
                     &                 & (sec.)                 & (- 497.0)     &                             \\
\hline
25 Mar 10 & 13:50:14 & 4$\times$800 & 0.674          & 10144                \\
                    & 18:57:49 & 4$\times$800 & 0.772          & 6921                   \\
26 Mar 10 & 14:08:19 & 4$\times$800 & 1.138         & 9113                    \\
                    & 17:03:17 & 4$\times$800 & 1.194         & 2034                    \\
                    & 19:17:19 & 4$\times$800 & 1.237         & 4127                    \\
27 Mar 10 & 14:57:05 & 4$\times$800 & 1.612         & 8385                    \\
                   & 16:49:52 & 2$\times$800 & 1.648         & 4045                    \\
                   & 17:19:28 & 2$\times$800 & 1.657        & 1470                     \\
                   & 19:06:03 & 4$\times$800 & 1.691        & 8869                     \\
28 Mar 10  & 14:33:02 & 4$\times$800 & 2.063        & 8330                     \\
                     & 17:05:00 & 4$\times$800 & 2.111        & 5814                     \\
31 Mar 10  & 13:06:47 & 4$\times$800 & 3.410        & 5950                      \\
                    & 19:01:29 & 4$\times$800 & 3.523       & 9673                      \\
01 Apr 10  & 13:30:07 & 4$\times$800 & 3.876         & 7274                      \\
                    & 18:45:07 & 4$\times$800 &  3.976       & 11575                    \\
02 Apr 10  & 13:03:07 & 4$\times$800 &  4.326       & 9440                       \\
                    & 18:45:48 & 4$\times$800 &  4.435       & 8337                       \\
03 Apr 10  & 13:16:00 & 4$\times$800 &  4.788       & 8470                       \\
                    & 15:50:33 & 4$\times$800 &  4.838      & 9398                        \\
                    & 18:55:30 & 4$\times$800 &  4.896      & 4473                        \\
04 Apr 10  & 14:22:06 & 4$\times$800 &  5.268       & 7227                        \\
                    & 18:39:53 & 4$\times$800 & 5.350       & 5186                        \\
05 Apr 10  & 12:05:53 & 4$\times$800 &  5.683       & 4779                       \\
                    & 13:05:15 & 4$\times$800 &  5.702       & 5559                        \\ 
\hline                                                                                                                                                                                                                                                                                                                                                                                                                                                                        
\end{tabular}
\end{table*}

The SEMPOL spectropolarimeter involves a fibre feed from the Cassegrain focus of the AAT to UCLES, where the two polarisation states (in this case left- and right-hand circular polarisation) are outputted in two fibres to UCLES with both polarisation states being recorded simultaneously on the detector. The basic construction of the SEMPOL polarimeter consists of a quarter-wave plate (or quarter-wave Fresnel rhomb, depending on instrument setup) and a half-wave Fresnel rhomb. Spectropolarimetric observations in circularly polarised light (Stokes V) consist of a sequence of four exposures. Between the exposures the half-wave Fresnel rhomb is rotated between +45\degrees and -45\degree. Thus the polarisation in each output fibre is alternated resulting in the removal of spurious polarisation signals from the telescope and polarimeter (at least to a first order approximation). Due to throughput loses in the spectropolarimeter setup the signal-to-noise (S/N) of spectropolarimetric data is usually lower than that obtained for straight spectroscopic observations. Further information on the operation of the SEMPOL spectropolarimeter can be found in \citet{SemelM:1993} and \citet{DonatiJF:1997c, DonatiJF:2003}.

As each Stokes V observation consists of 4 exposures this means that we generally have 4 times the number of Stokes I (intensity) observations as we do Stokes V. Note however that due to poor weather during some runs there are a few sequences on HD 141943 where we were only able to obtain a sequence of two exposures (see Table~\ref{Tab_log}). The Stokes V profile can still be extracted from just two exposures, but with typically lower S/N.

The same detector and wavelength setup was used for all 4 observing runs. The detector was the EEV2 CCD with 2048 $\times$ 4096 13.5 $\mu$m square pixels. As the EEV2 is larger than the unvignetted field of UCLES a smaller window format (2048 $\times$ 2746 pixels) was used to reduce readout time. Using the 31 gr/mm grating, 46 orders (\#129 to \#84) were observed, giving a full wavelength coverage from $\sim$4380 \AA\/ to $\sim$6810 \AA. For the May 2006 observations the chip was binned by 2 in the spectral direction and a 1 arcsec slit was used providing a resolution of $\sim$50,000 (i.e.\ $\sim$6.0 \kmsn). For the spectropolarimetric observations (in 2007, 2009 and 2010) there was no binning and the fibre feed provided a resolution of $\sim$70,000 (i.e.\ $\sim$4.3 \kmsn).

All raw frames were reduced into wavelength calibrated spectra using the ESpRIT (Echelle Spectra Reduction: an Interactive Tool) optimal extraction routines of \citet{DonatiJF:1997c}. As the Zeeman
signatures in atomic lines are extremely small (typical relative amplitudes of 0.1 per cent or less) we have applied the technique of Least-Squares Deconvolution \citep[LSD,][]{DonatiJF:1997c} to the over 2600 photospheric spectral lines in each echelle spectrum in order to create a single high S/N profile for each observation. The line mask used for the LSD was a G2 line list created from the Kurucz atomic database and ATLAS9 atmospheric models \citep{KuruczRL:1993}. 

The peak S/N of the initial Stokes I observations ranged from 30 to 260 (depending on observing conditions) while the peak S/N for the Stokes V observations (combining 2 or 4 exposures) was 50 to 330. LSD has been applied to both the Stokes I and Stokes V data resulting in S/N values of 540 to 930 for the Stokes I profiles and 1450 to 11600 for the Stokes V profiles. This corresponds to a multiplex gain of $\sim$4 -- 10 for the Stokes I observations and $\sim$35 for the Stokes V observations. The multiplex gain for the Stokes I profiles is significantly less than that for the Stokes V profiles because, as pointed out by \citet{DonatiJF:1997c}, the technique of LSD appears to be not as suited to Stokes I as it is to Stokes V. In addition, the calculated S/N of the Stokes I LSD profiles are usually somewhat underestimated (see Section~\ref{Sec_bri}). Further information on LSD can be found in \citet{DonatiJF:1997c} and \citet*{KochukhovO:2010}.

In order to correct for wavelength shifts of instrumental origin, each spectrum was shifted to match the Stokes I LSD profile of the telluric lines contained in the spectrum, as has been done by \citet{DonatiJF:2003} and other authors. This reduces the relative radial velocity shifts of the LSD profiles to $\pm$ $\sim$0.1 \kmsn.

\section{Fundamental parameters of HD 141943} \label{Sec_fun}

In order to determine the surface topologies of HD 141943 the ZDI code needs to know several basic stellar parameters, namely the \vsini, radial velocity, inclination of the rotation axis to the observer, and the temperature of the photosphere and spots on the stellar surface. The \vsinis and radial velocity are determined using the $\chi^{2}$-minimisation technique of \cite{BarnesJR:2000}. This is simply using those parameters that give the best fit to the dataset (lowest reduced $\chi^{2}$ values).

According to \citet{HillenbrandLA:2008} HD 141943 is at a distance of 67 pc, has an effective temperature of 5805 K and a luminosity of 2.7 L\subs{\odot}. Now these latter two values will be slightly higher for an unspotted star, but we do not know the spottedness of the star when the \citet{HillenbrandLA:2008} determinations were made. Given the level of spottedness shown on HD 141943 (see Section~\ref{Sec_bri}) we have assumed a spot coverage of 0\% to 5\% with the most likely coverage being 3\%. This assumption, along with a spot-photosphere temperature difference of $\sim$1900 K \citep[calculated from Fig. 7 of][]{BerdyuginaSV:2005}, gives a photospheric temperature of $\sim$5850 K, a spot temperature of $\sim$3950 K and an unspotted luminosity of $\sim$2.8  L\subs{\odot}.

As there are no error measurements given for the \citet{HillenbrandLA:2008} measurements we have assumed an error of $\pm$ 100 K in the photospheric temperature and $\pm$ 0.1 L\subs{\odot} in the luminosity. These, along with the uncertain spot coverage at the time of the \citet{HillenbrandLA:2008} determinations and a \vsinis of 35 $\pm$ 0.5 \kmsn, imply that HD 141943 has a radius of $\sim$1.6 $\pm$ 0.15 R\subs{\odot} and an inclination angle of $\sim$70\sups{\circ}$^{+20^{\circ}}_{-10^{\circ}}$.  With a high inclination angle our imaging code (see Section~\ref{Sec_ima}) has difficulty determining if features are located in the Northern or Southern hemisphere so we have limited the inclination angle range to 70\sups{\circ} $\pm$ 10\sups{\circ}. Table~\ref{Tab_param} lists the basic stellar parameters of HD 141943 that we have determined for this study. As mentioned, there are no error measurements given for the \citet{HillenbrandLA:2008} measurements of effective temperature and luminosity, so the estimations of the errors in the stellar parameters of HD 141943 may well be underestimates.

\begin{figure}
  \centering
  \includegraphics[angle=90, width=\columnwidth]{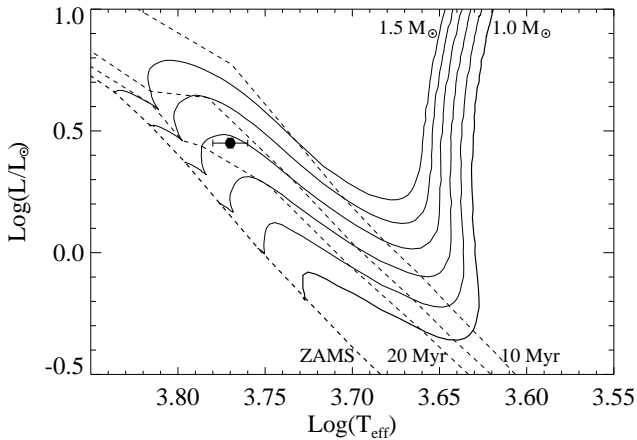}
  \caption{The evolutionary status of HD 141943 from the pre-main sequence models of \citet{SiessL:2000}, using the 2007 updates. The solid lines are pre-main sequence evolutionary tracks for stellar masses of 1.0 M\subs{\odot} to 1.5 M\subs{\odot} in 0.1 M\subs{\odot} steps, with the bold solid line denoting the 1.0 M\subs{\odot} track. The dashed lines are age isochrones for stellar ages of 10 Myr, 15 Myr, 20 Myr, and the Zero-Age Main-Sequence (bold dashed line).} 
  \label{Fig_Siess}
\end{figure}

From the values for photospheric temperature and unspotted luminosity we can determine the mass and age of HD 141943 from the evolutionary models of \citet*{SiessL:2000}, see Fig.~\ref{Fig_Siess}. This makes HD 141943 a $\sim$1.3 M\subs{\odot} star of age $\sim$17 Myrs, in reasonable agreement with the values from \citet{CutispotoG:2003}, and thus it can be classified as a PMS star.

\begin{table}
\caption{Fundamental parameters of HD 141943 used/found in this study. The age and mass are from the theoretical isochrones of \citet{SiessL:2000} given in Fig.~\ref{Fig_Siess} and the equatorial rotation period has been determined from the differential rotation, see Paper II}
\label{Tab_param}
\centering
\begin{tabular}{lc}
\hline\hline
Parameter & value\\
\hline
Age & $\sim$17 Myrs\\
Mass & $\sim$1.3 M\subs{\odot}\\
Photospheric temperature & 5850 $\pm$ 100 K$^{a}$\\
Spot temperature & $\sim$3950 K\\
Unspotted luminosity & 2.8 $\pm$ 0.1 L$^{a}_{\odot}$\\
Stellar radius & 1.6 $\pm$ 0.15 R\subs{\odot}\\
\vsinis & 35.0 $\pm$ 0.5 \kmss\\
Radial velocity ($v_{\rm rad}$) & see Table~\ref{Tab_radvel}\\
Inclination angle ($i$) & 70\sups{\circ} $\pm$ 10\sups{\circ}\\
Equatorial rotation period ($P_{\rm eq}$) & $\sim$2.182 days\\
\hline
$^{a}$assumed error.
\end{tabular}
\end{table}

\subsection{Is HD 141943 a binary?}\label{Sec_bin}

\citet{NordstromB:2004} took 7 radial velocity measurements of HD 141943 over a 2979 day period (epochs and measurements were not given). Based on the fact that the standard error of each measurement was less than the standard deviation of the measurements they determined that HD 141943 only has a 31.4 per cent chance of having a constant radial velocity. 

Our own radial velocity measurements also show some level of variation. Taking the radial velocity measurement of \citet{WaiteIA:2005} along with ours (\citealt{CutispotoG:1999} has some radial velocity measurements but the errors are too large to be useful) we have listed them in Table~\ref{Tab_radvel} to highlight the level of variation. The radial velocity for the 2006, 2007, 2009 and 2010 datasets has been calculated as part of the Doppler imaging process (see Section~\ref{Sec_ima}) and are for the entire dataset as a whole. Given that the radial velocity for 2007, 2009 and 2010 is constant (within error bars) it would mean that if HD 141943 is a binary then it would appear to have an elliptical orbit. 

\begin{table}
\caption{Log of radial velocity measurements of HD 141943 from these observations plus that of \citet{WaiteIA:2005}. For the 2006, 2007, 2009 and 2010 datasets the HJD given is for the mid-point of the observations. The errors in the radial velocity measures represent a 3$\sigma$ change in the reduced $\chi^{2}$ values (see text). }
\label{Tab_radvel}
\centering 
\begin{tabular}{ccc}
\hline\hline
Data set & HJD & Radial Velocity\\
 & (2450000.0+) & (\kmsn)\\
\hline
2005 & 2396.828 & -1.5 $\pm$ 1.0\\
2006 & 3864.801 & -0.6 $\pm$ 0.1\\
2007 & 4195.154 & 0.1 $\pm$ 0.1\\
2009 & 4932.071 & 0.1 $\pm$ 0.1\\
2010 & 5286.564 & 0.2 $\pm$ 0.1\\
\hline
\end{tabular}
\end{table}

\section{Images of HD 141943} \label{Sec_ima}

A surface image of a rapidly-rotating star can be generated through the inversion of a time series of Stokes I (brightness images) or Stokes V (magnetic images) profiles.  The brightness and magnetic topologies of HD 141943 were created using the ZDI code of \citet{BrownSF:1991} and \citet{DonatiJF:1997b} using a linear limb-darkening coefficient of 0.66. As the inversion is an ill-posed problem, an infinite number of solutions can be found that fit the data to the noise level. Therefore a regularisation scheme is usually introduced to determine a unique solution with the role of the regularisation scheme being reduced with high quality data. Our imaging code uses the \citet{SkillingJ:1984} maximum-entropy regularisation scheme. This scheme has the effect of producing images with the minimum amount of information required to fit the data to the noise level.

\subsection{Brightness images} \label{Sec_bri}

The imaging code to reconstruct the spot features on HD 141943 uses a two-temperature model (one for the spots and the other for the unspotted photosphere) as described by \citet{CameronAC:1992}. Each pixel on the stellar surface is reconstructed for spot occupancy (the local relative area occupied by cool spots) and varies from 0 (no spots) to 1 (maximum spottedness). Following on from \citet{UnruhYC:1995}, who showed that there is little change in spot topology when using a synthetic profile over profiles created from slowly rotating comparison stars, we have used Gaussian profiles to represent the profiles of both the spots and the photosphere. This is almost standard practice now with a number of previous works doing the same \citep*[i.e.][]{PetitP:2004b, MarsdenSC:2005b, MarsdenSC:2006a, JeffersSV:2008}. As has done for the two previous young early-G stars for which spot topologies have been obtained by this method (R58, \citealt{MarsdenSC:2005b} and HD 171488, \citealt{MarsdenSC:2006a}) we have used the same Gaussian to represent both the spot and photosphere. The full-width at half-maximum (FWHM) of the Gaussians used were slightly different for the spectroscopic and spectropolarimetric data sets, with a FWHM of 10 \kmss used for the spectroscopic data set and a FWHM of 9 \kmss used for the spectropolarimetric data sets. These FWHM were determined from matching the FWHM of the Moon's LSD profile taken with the same instrumental setup during each observing run.

\begin{figure}
  \centering
  \includegraphics[angle=-90, width=\columnwidth]{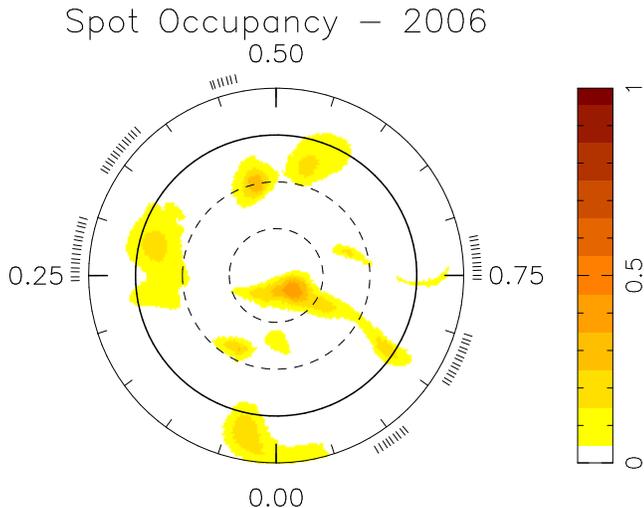}
  \caption{Maximum entropy brightness image reconstruction for HD 141943, May 2006. The image is a flattened polar projection looking down on the north pole and extending down to -30\degrees latitude. The bold line denotes the equator and the dashed lines are +30\degrees and +60\degrees latitude parallels. The radial ticks outside the plot indicate the phases at which the star was observed. The image has a spot filling factor of 0.021 (or 2.1 per cent). The image incorporates the measured surface differential rotation, see Paper II.} 
  \label{Fig_map2006}
\end{figure}

\begin{figure*}
  \centering
  \includegraphics[angle=-90, width=\textwidth]{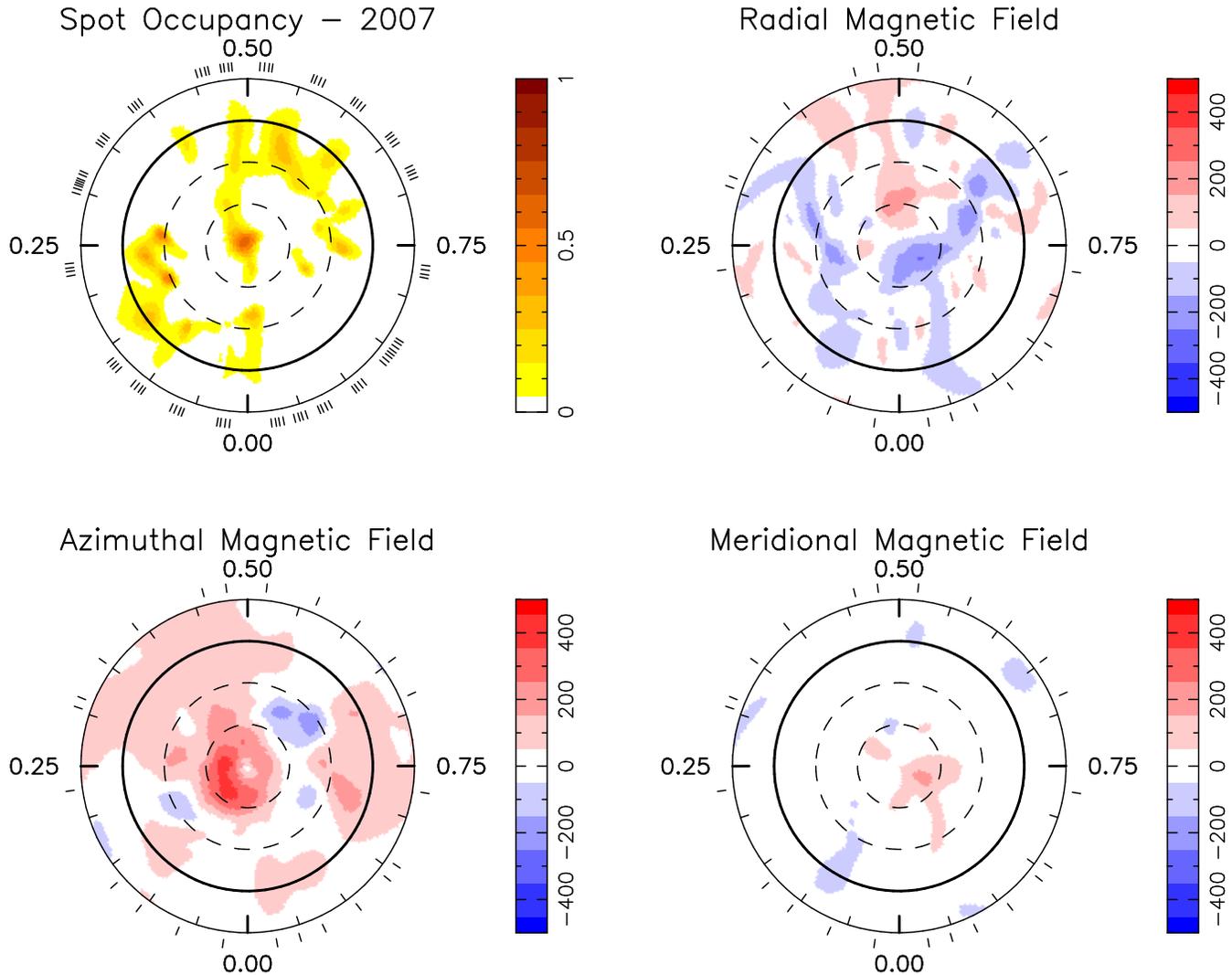}
  \caption{Maximum entropy brightness and magnetic image reconstructions for HD 141943, March/April 2007. The images are flattened polar projections as described in Fig.~\ref{Fig_map2006}. The scale in the magnetic images is in Gauss. The brightness image (top-left) has a spot filling factor of 0.031 (or 3.1 per cent), while the magnetic images have a mean field modulus of 91.3 G. The images have been created with the inclusion of the surface differential rotation of the star, see Paper II.} 
  \label{Fig_allmap2007}
\end{figure*}

\begin{figure*}
  \centering
  \includegraphics[angle=-90, width=\textwidth]{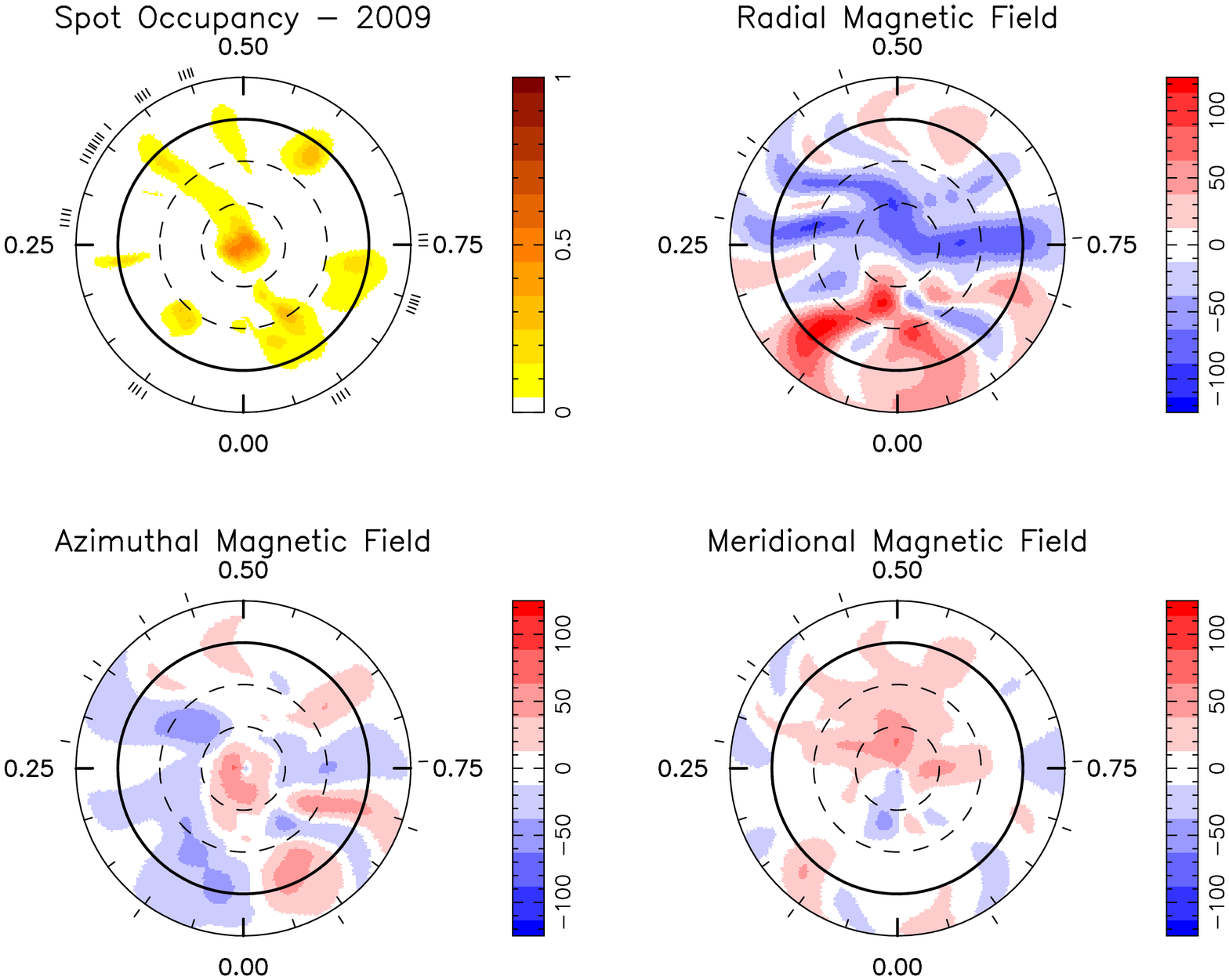}
  \caption{Maximum entropy brightness and magnetic image reconstructions for HD 141943, April 2009. The images are flattened polar projections as described in Fig.~\ref{Fig_map2006}. The brightness image (top-left) has a spot filling factor of 0.027 (or 2.7 per cent), while the magnetic images have a mean field modulus of 36.6 G. The images incorporate the measured surface differential rotation, see Paper II. Note the different scale in the magnetic images compared to Fig.~\ref{Fig_allmap2007}.} 
  \label{Fig_allmap2009}
\end{figure*}

\begin{figure*}
  \centering
  \includegraphics[angle=-90, width=\textwidth]{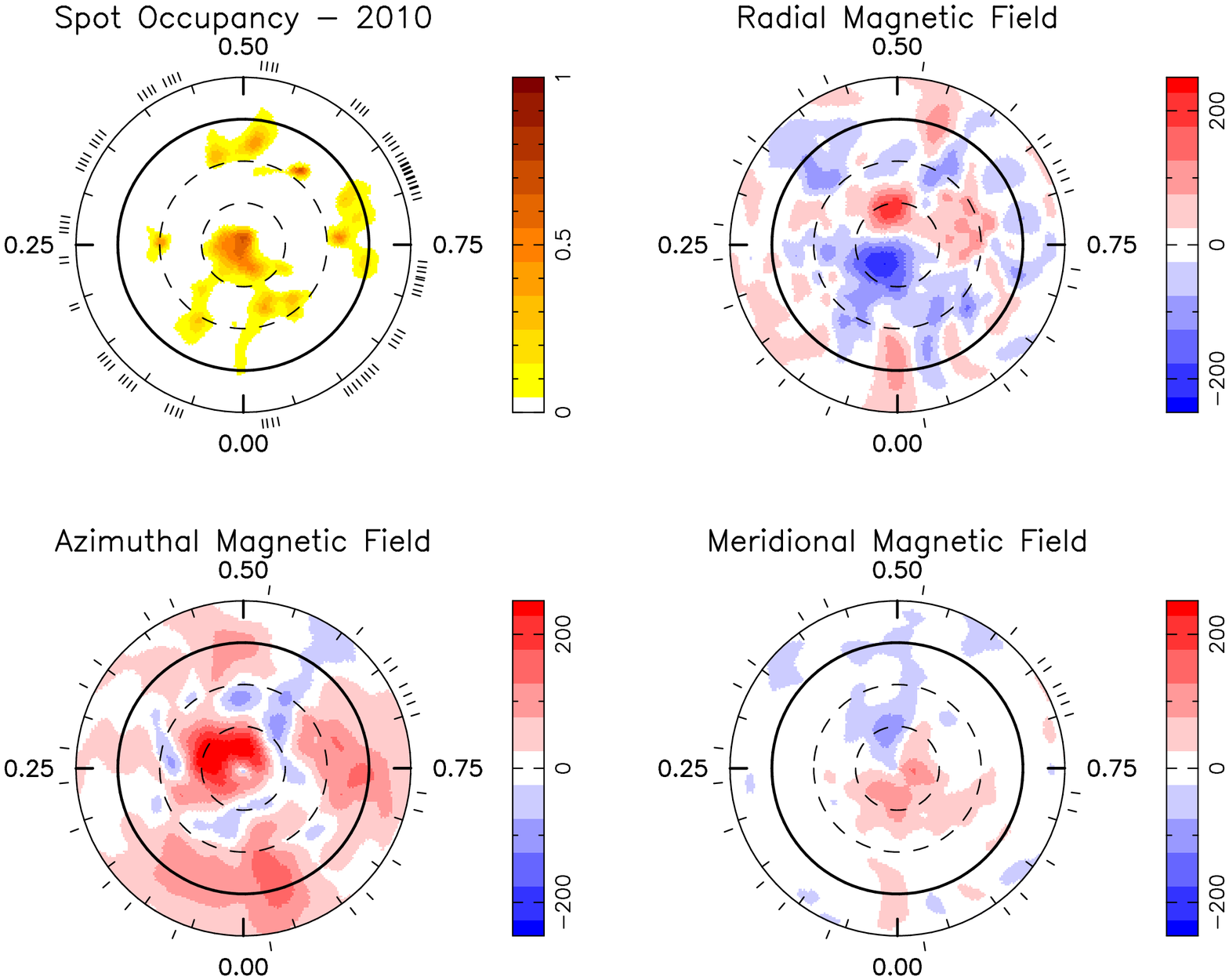}
  \caption{Maximum entropy brightness and magnetic image reconstructions for HD 141943, March/April 2010. The images are flattened polar projections as described in Fig.~\ref{Fig_map2006}. The brightness image (top-left) has a spot filling factor of 0.029 (or 2.9 per cent), while the magnetic images have a mean field modulus of 70.9 G. The images incorporate the measured surface differential rotation, see Paper II. Note the different scale in the magnetic images compared to Fig.~\ref{Fig_allmap2007} and Fig~\ref{Fig_allmap2009}.} 
  \label{Fig_allmap2010}
\end{figure*}

\begin{figure*}
  \centering
  \includegraphics[angle=90, width=\textwidth]{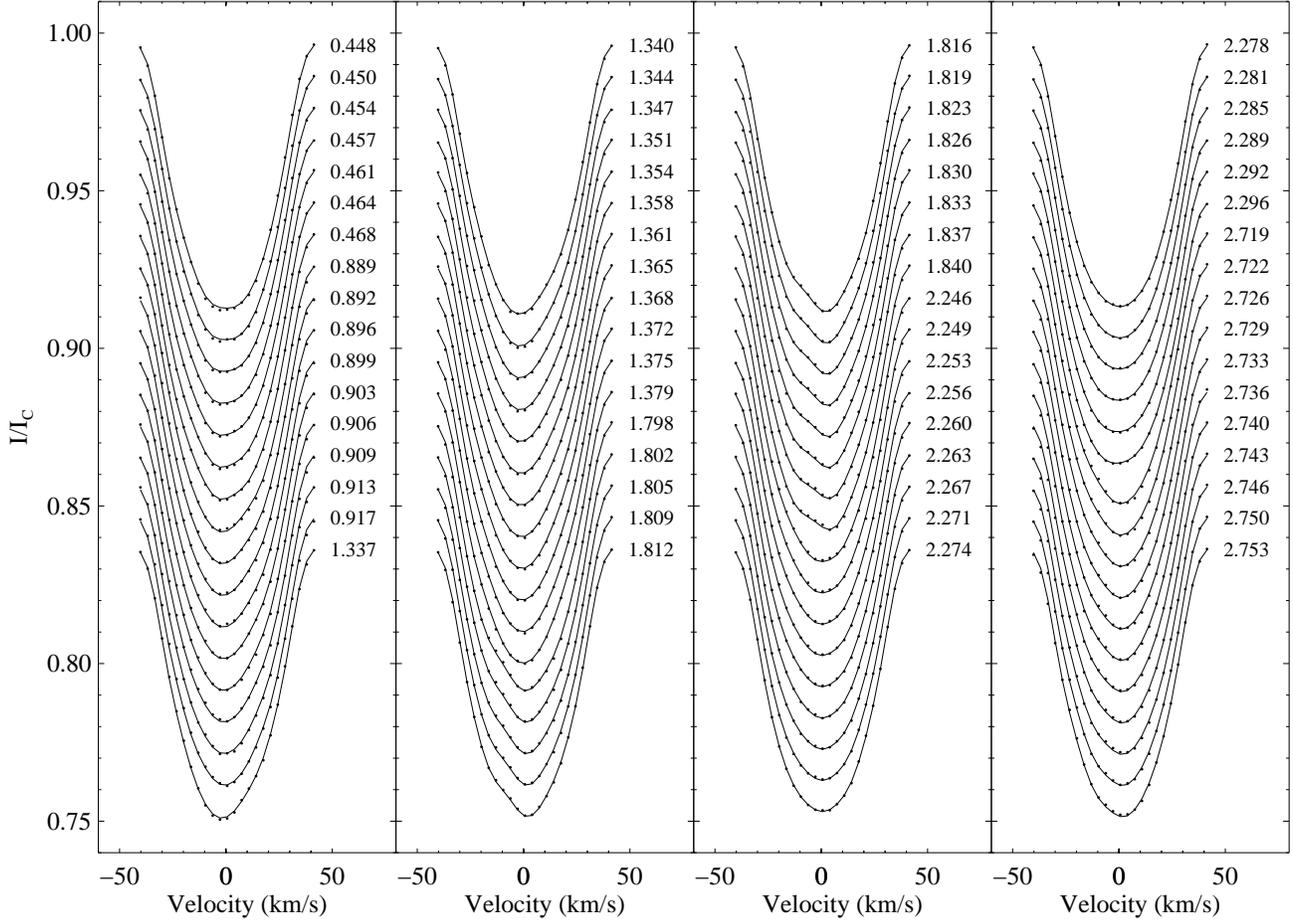}
  \caption{Maximum entropy fits for the Stokes I LSD profiles of HD 141943, May 2006. The dots represent the observed LSD profiles while the lines represent the fits to the profiles produced by the imaging code. Each profile is shifted down by 0.01 for graphical purposes. The rotational phases at which the observations took place are indicated to the right of each profile.}
  \label{Fig_brifit2006}
\end{figure*}

\begin{figure*}
  \centering
  \includegraphics[angle=90, width=\textwidth]{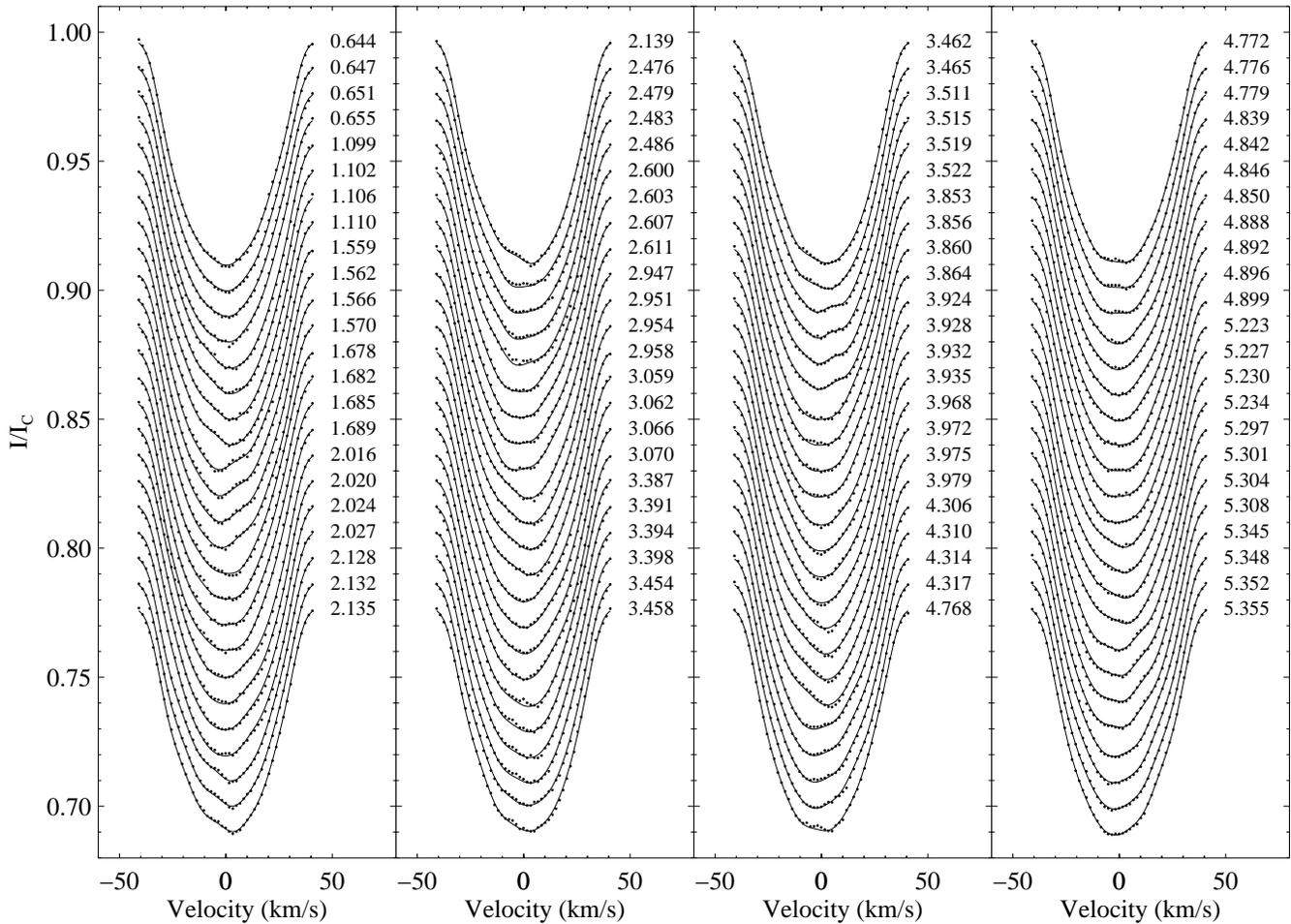}
  \caption{Maximum entropy fits for the Stokes I LSD profiles of HD 141943, March/April 2007. Again the dots represent the observed LSD profiles while the lines represent the fits to the profiles produced by the imaging code as described in Fig.~\ref{Fig_brifit2006}.}
  \label{Fig_brifit2007}
\end{figure*}

\begin{figure}
  \centering
  \includegraphics[width=\columnwidth]{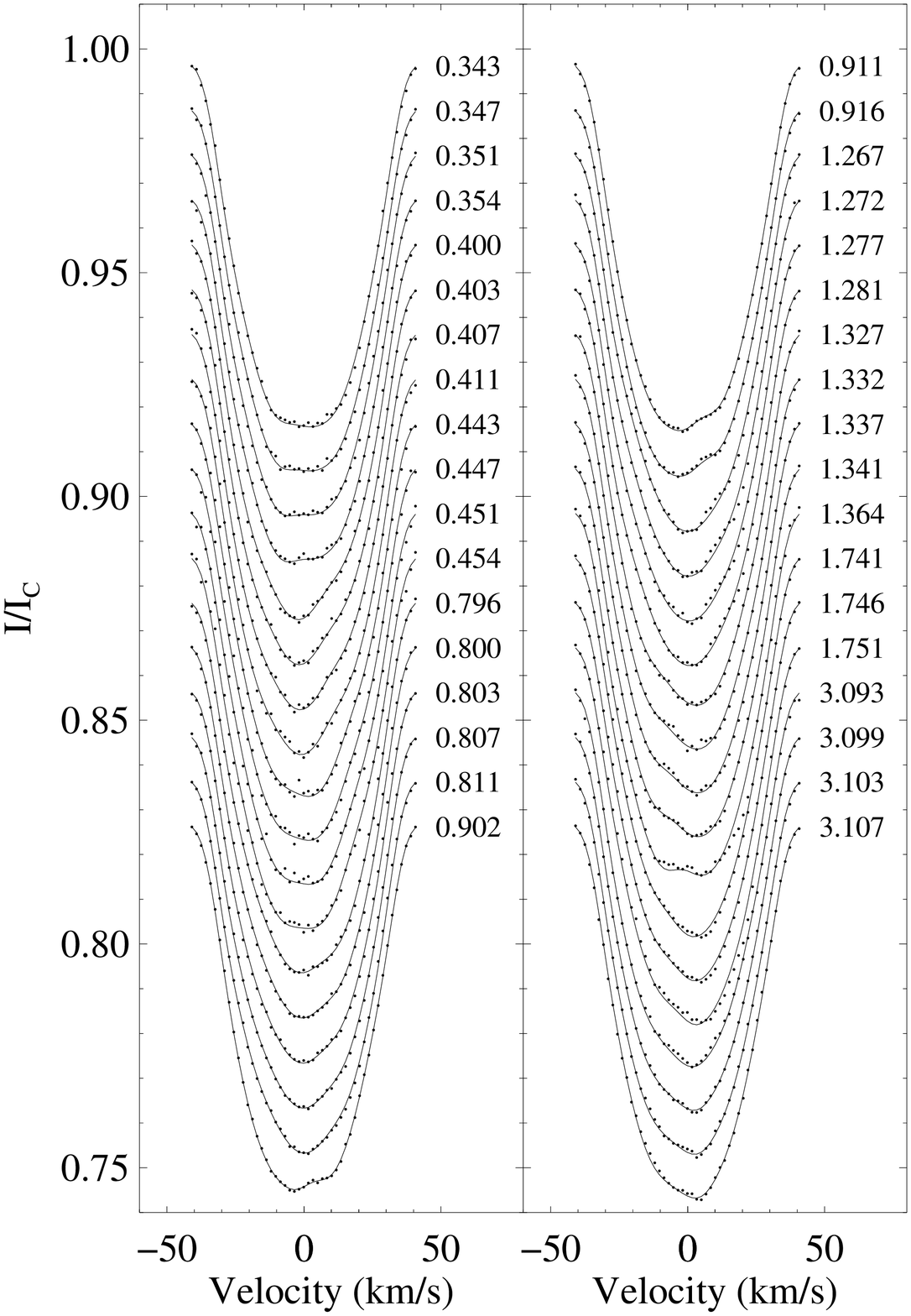}
  \caption{Maximum entropy fits for the Stokes I LSD profiles of HD 141943, April 2009. As described in Fig.~\ref{Fig_brifit2006}, the dots represent the observed LSD profiles while the lines represent the fits to the profiles produced by the imaging code.}
  \label{Fig_brifit2009}
\end{figure}

\begin{figure*}
  \centering
  \includegraphics[angle=90, width=\textwidth]{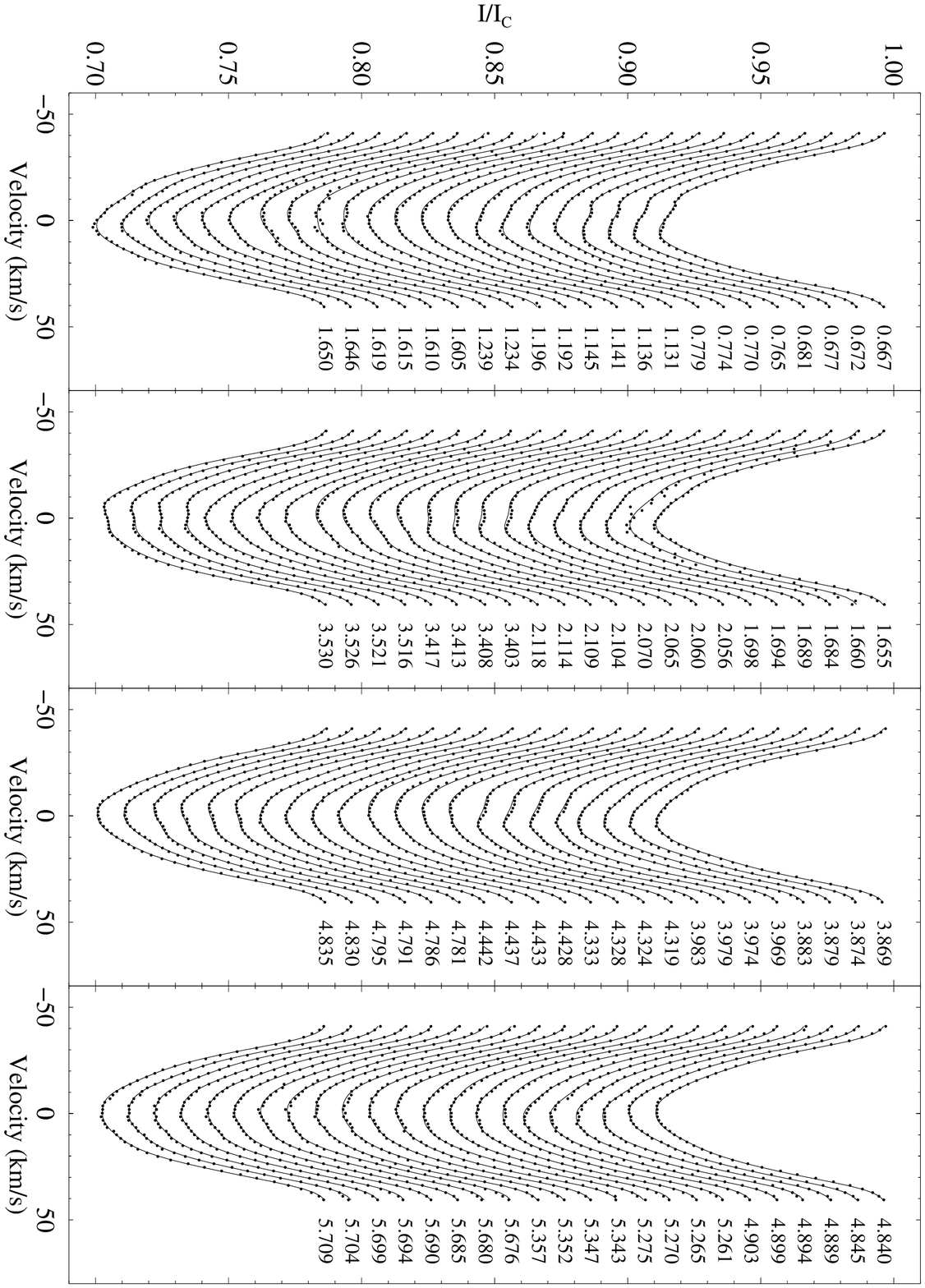}
  \caption{Maximum entropy fits for the Stokes I LSD profiles of HD 141943, March/April 2010. As described in Fig.~\ref{Fig_brifit2006}, the dots represent the observed LSD profiles while the lines represent the fits to the profiles produced by the imaging code.}
  \label{Fig_brifit2010}
\end{figure*}

The maximum-entropy brightness images for HD 141943 for the 4 observing epochs are shown in Fig.~\ref{Fig_map2006} (2006 observations), the top-left image of Fig.~\ref{Fig_allmap2007} (2007 observations), the top-left image of Fig.~\ref{Fig_allmap2009} (2009 observations) and the top-left image of Fig~\ref{Fig_allmap2010} (2010 observations). These images were created fitting the data down to reduced $\chi^{2}$ values of 0.2, 0.3, 0.4 and 0.3 for the 2006, 2007, 2009 and 2010 datasets respectively. As explained in \citet{PetitP:2004b} a reduced $\chi^{2}$ value smaller than unity can be achieved because, as mentioned in Section~\ref{Sec_obs}, the S/N calculated for the Stokes I LSD profiles are underestimated. This has no effect on the maps produced. Fits of the modelled profiles to the observed LSD profiles are given in Fig.~\ref{Fig_brifit2006} (2006 observations), Fig.~\ref{Fig_brifit2007} (2007 observations), Fig.~\ref{Fig_brifit2009} (2009 observations) and Fig~\ref{Fig_brifit2010} (2010 observations). The epochs of the observations, as calculated from the midpoint of each dataset, are: 2006.352, 2007.257, 2009.273 and 2010.244 in decimal years.

The brightness images of HD 141943 show that at all 4 epochs HD 141943 had a smallish polar spot and numerous low-latitude features situated mostly between the equator and $\sim$+30\sups{\circ} latitude. In the May 2006 image (Fig.~\ref{Fig_map2006}) there appears to be some features below the equator and above +30\sups{\circ}. However, as there are no observations around these phases it could well be that the code has had trouble determining the latitude of spot features that are only seen in the wings of the profiles. 

The spot filling factor (the level of spot coverage over the entire stellar surface) is also very similar over the 4 epochs. In 2006 the spot coverage was 2.1 per cent, in 2007 it was 3.1 per cent, in 2009 2.7 per cent and in 2010 it was 2.9 per cent. The reason for the slightly lower value of spot coverage for the 2006 dataset may well be due to the limited phase coverage of the observations, see Fig.~\ref{Fig_map2006}. Thus some spots in these observing gaps may not have been recovered.

\begin{figure}
  \centering
  \includegraphics[angle=90, width=\columnwidth]{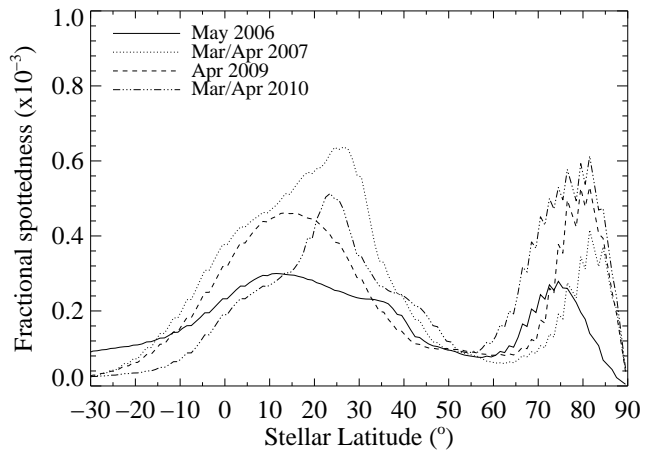}
  \caption{Fractional spottedness versus stellar latitude for HD141943. Fractional spottedness is based on the average spot occupancy at each latitude and is defined by equation~\ref{Eqn_frac}.}
  \label{Fig_frac_spot_lat}
\end{figure}

The variation in spot occupancy with stellar latitude is given in Fig.~\ref{Fig_frac_spot_lat}, which plots fractional spottedness versus stellar latitude. Fractional spottedness is defined as:
\BE
F(\theta) = \frac{S(\theta)cos(\theta)d\theta}{2} \label{Eqn_frac}
\EE
where, $S(\theta)$ is the average spot occupancy at latitude $\theta$ and $d\theta$ is the latitude width of each latitude ring.

\begin{figure}
  \centering
  \includegraphics[width=\columnwidth]{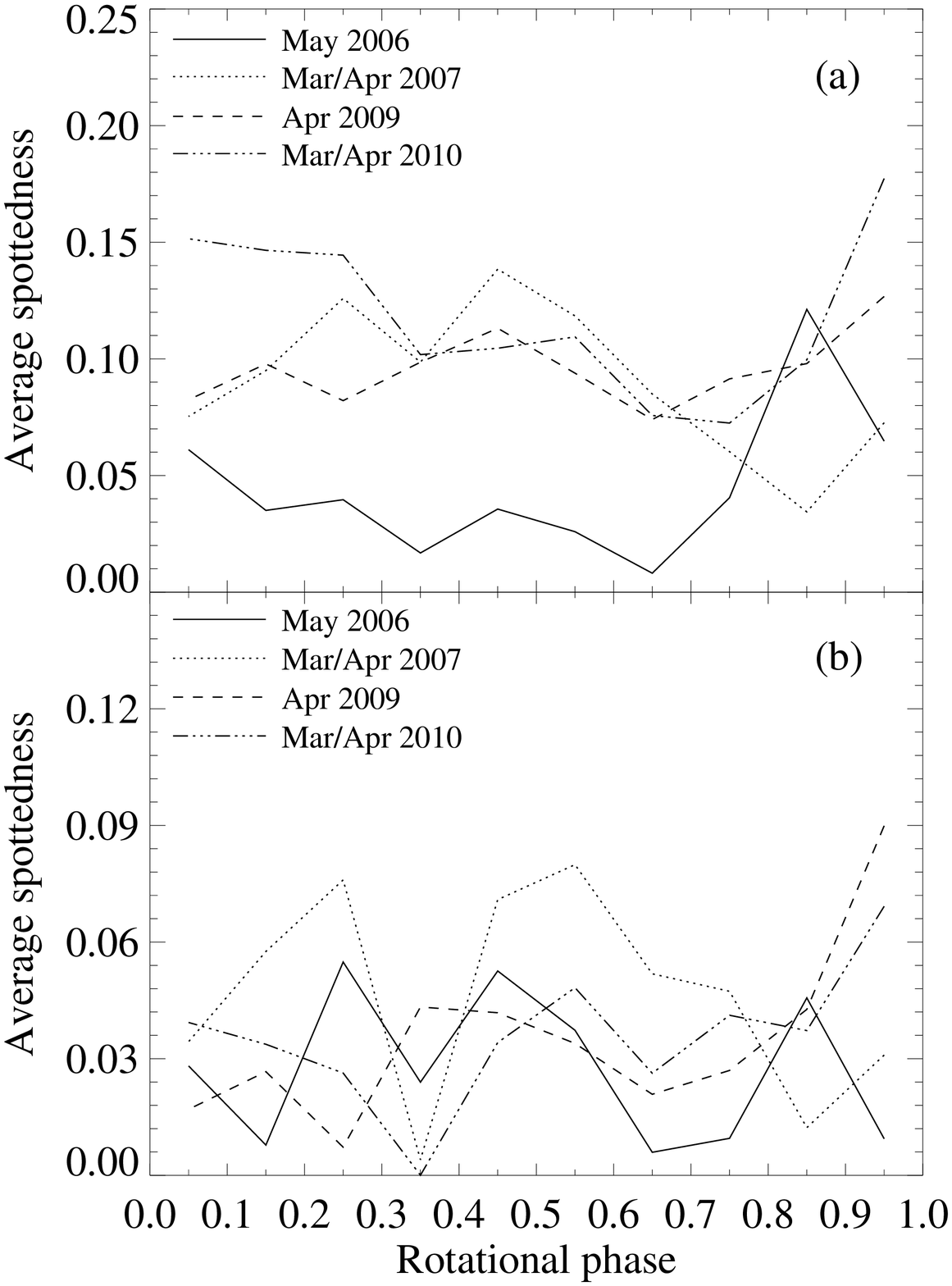}
  \caption{Average spottedness versus stellar rotational phase for HD141943, averaged over (a) 0\sups{\circ} to +90\sups{\circ} and (b) 0\sups{\circ} to +60\sups{\circ} latitude. The phase has been binned into 0.1 steps to remove any small scale variations.}
  \label{Fig_frac_spot_lng}
\end{figure}

It has been reported that active solar-type stars show evidence of active longitudes, where certain longitudes (usually two longitudes around 180\sups{\circ} apart) are more active than others on the stellar surface \citep[i.e.][]{BerdyuginaSV:1998, JarvinenSP:2005}. In order to test this we have plotted the average spottedness versus stellar rotational phase for all 4 epochs of HD 141943 observations. This is displayed in Fig.~\ref{Fig_frac_spot_lng}. In case the polar spot is affecting the results we have plotted the average spottedness for both (a) 0\sups{\circ} to +90\sups{\circ} and (b) 0\sups{\circ} to +60\sups{\circ} in latitude.

For the 2006 and 2009 datasets there is very little evidence for active longitudes on the stellar surface and the limited phases observed in these datasets may well be influencing this result. For the well sampled 2007 dataset, and looking at the top-left image in Fig.~\ref{Fig_allmap2007}, there appears to be a slight enhancement of the lower-latitude spots around phases $\sim$0.25 and $\sim$0.55. However as shown in Fig.~\ref{Fig_frac_spot_lng}(b) these enhancements are rather broad. For the 2010 dataset there appears to be an increase in the spot coverage at phase $\sim$0.00. However, this is reduced when only considering low-latitude features (Fig.~\ref{Fig_frac_spot_lng}(b)) and when looking at the top-left image in Fig.~\ref{Fig_allmap2010} the enhancement could be caused by the extension of the polar spot to slightly lower-latitudes around phase $\sim$0.00 in the 2010 epoch. There would not appear to be any evidence of another active longitude on the star in 2010. We do not feel that these represent strong evidence for active longitudes on HD 141943.

\subsection{Magnetic images} \label{Sec_mag}

The magnetic field topology for HD 141943 has been reconstructed for three epochs using the modelling of \citet{DonatiJF:1997b} and including the spherical harmonic expansion of the surface magnetic field by \citet{DonatiJF:2006}. The magnetic reconstruction is done assuming a generalised potential field plus a toroidal field using a high order ($l$ $\le$ 30) spherical harmonic expansion. A limit of $l$\subs{max} $=$ 30 was chosen for the spherical harmonic expansion, as beyond this limit the magnetic topologies remain essentially unchanged.

Various weighting schemes can be applied to the spherical harmonic expansion so that the reconstruction favours different magnetic field topologies. Following the principle of Occam's razor, we have used a weighting scheme that favours ``simpler'' magnetic fields (i.e.\ those with lower $l$ values) while still reconstructing a similar overall magnetic topology to an unweighted reconstruction. Details of the spherical harmonic technique can be found in \citet{DonatiJF:2006}.

It should be noted that the April 2009 dataset can be modelled using a purely potential field without the need to include a toroidal field, but this may be due to the limited data in the 2009 observations. The March/April 2007 and March/April 2010 datasets however, cannot be modelled to a reduced $\chi^{2}$ value of 1.0 without the inclusion of a toroidal field component. Thus we have assumed a potential field plus toroidal field when modelling all three datasets.

Using the Stokes V LSD profiles the magnetic imaging code reconstructs images of radial, azimuthal and meridional field on the stellar surface, assuming a weak magnetic field and a constant Gaussian profile over the stellar surface \citep[see][]{DonatiJF:2003}. Radial field is defined as field in/out of the stellar surface, with positive field being field lines directed outward from the surface. Azimuthal field is defined as field wrapped around the rotational axis of the star, with positive being defined as counterclockwise. Meridional field is defined as field following lines of longitude north and south, with positive being northward.

\begin{figure}
  \centering
  \includegraphics[width=\columnwidth]{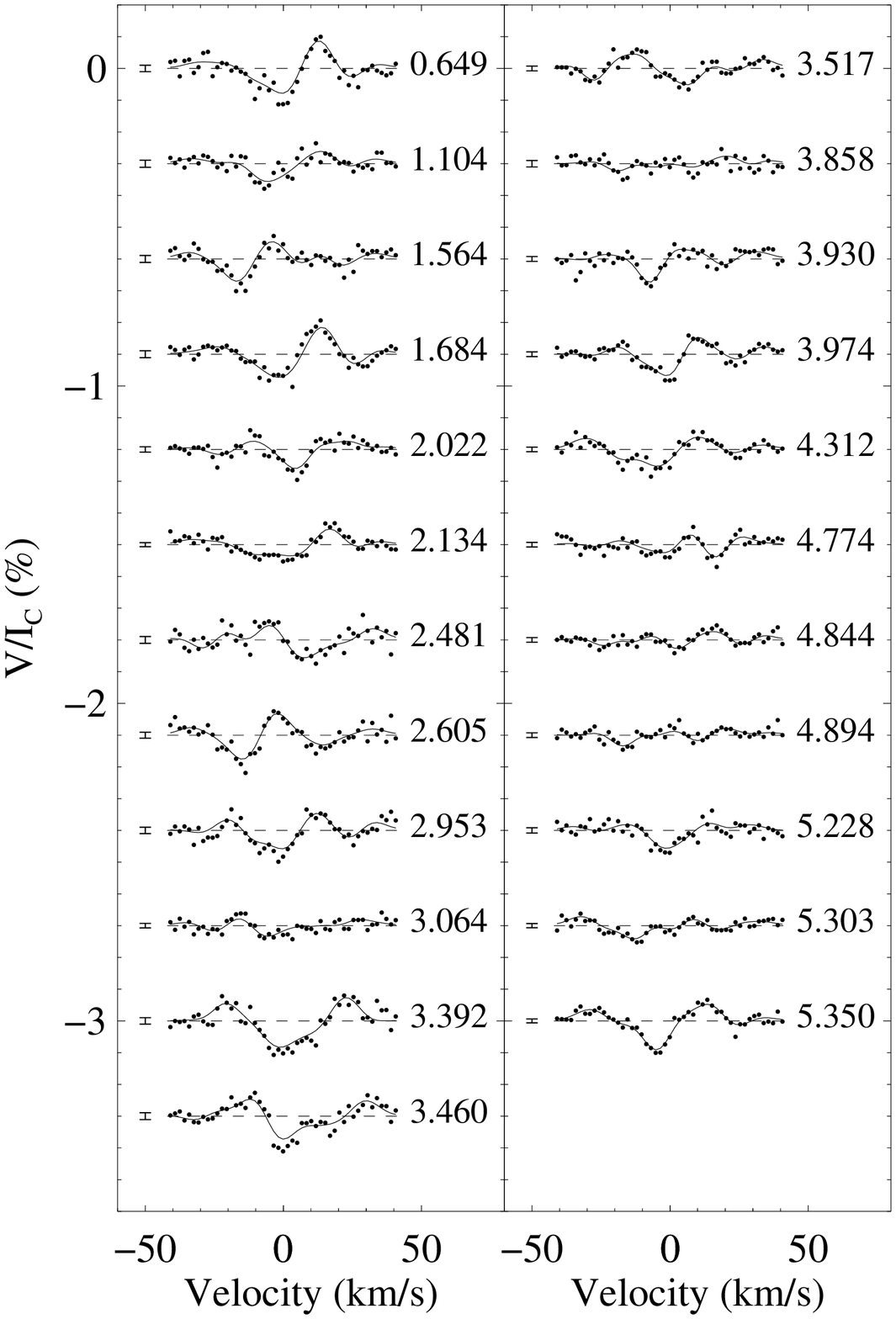}
  \caption{Maximum entropy fits for the Stokes V LSD profiles of HD 141943, March/April 2007. The dots represent the observed LSD profiles while the lines represent the fits to the profiles produced by the imaging code. Each profile is shifted down by 0.003 for graphical purposes. The rotational phases at which the observations took place are indicated to the right of each profile while the 1$\sigma$ error bars are given to the left of each profile.}
  \label{Fig_magfit2007}
\end{figure}

\begin{figure}
  \centering
  \includegraphics[width=0.61\columnwidth]{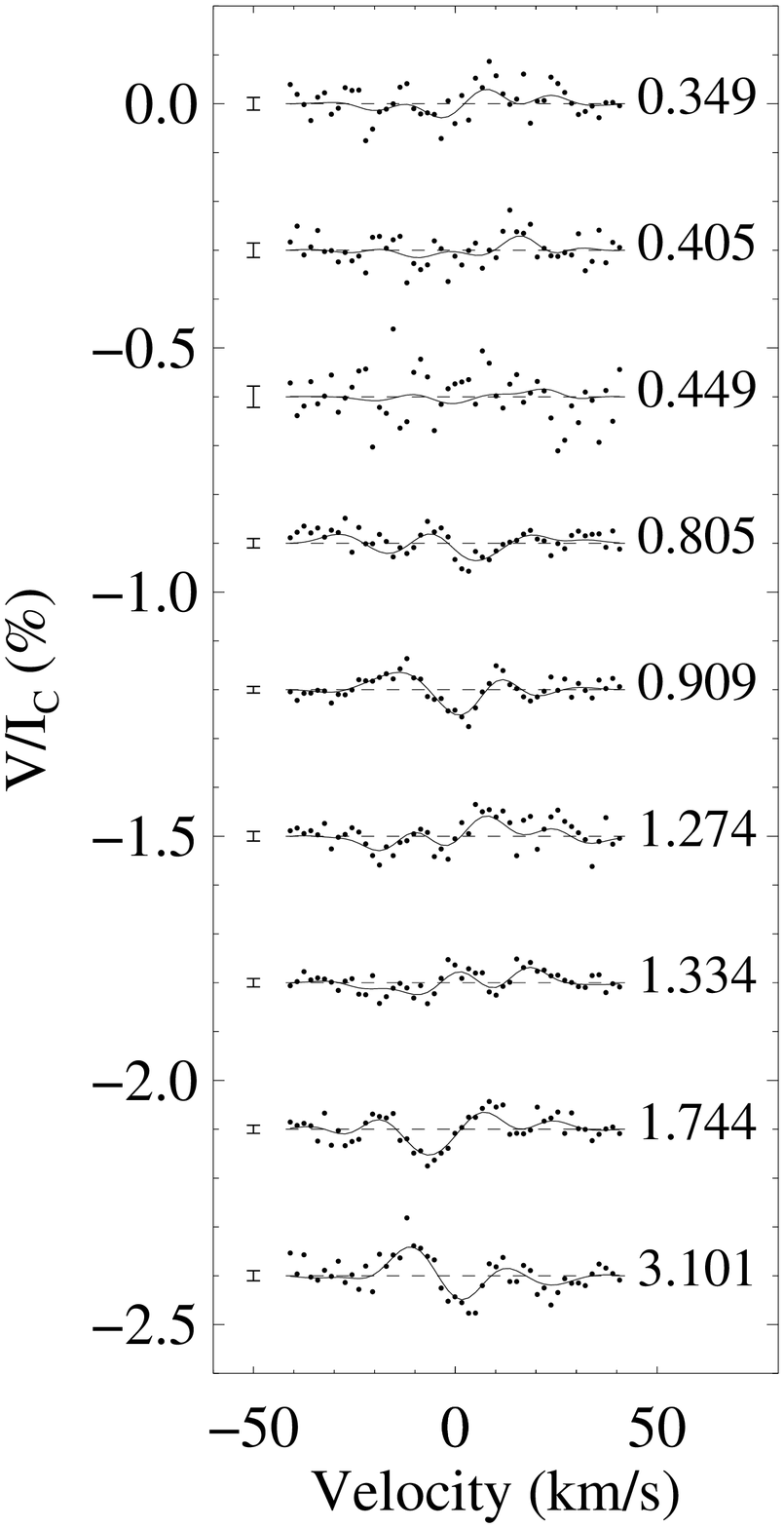}
  \caption{Maximum entropy fits for the Stokes V LSD profiles of HD 141943, April 2009. As explained in Fig.~\ref{Fig_magfit2007}, the dots represent the observed LSD profiles while the lines represent the fits to the profiles produced by the imaging code.}
  \label{Fig_magfit2009}
\end{figure}

\begin{figure}
  \centering
  \includegraphics[width=\columnwidth]{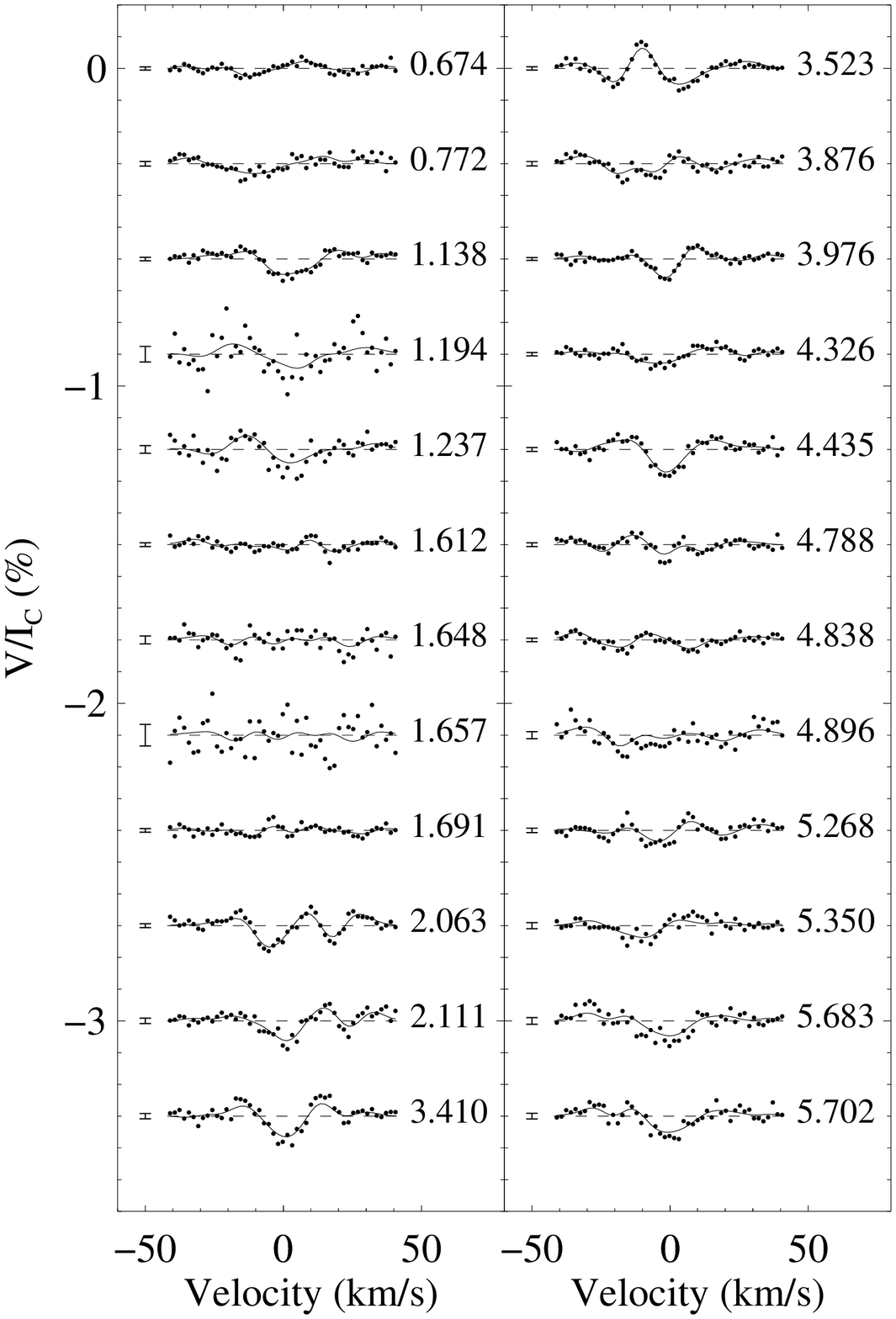}
  \caption{Maximum entropy fits for the Stokes V LSD profiles of HD 141943, March/April 2010. As explained in Fig.~\ref{Fig_magfit2007}, the dots represent the observed LSD profiles while the lines represent the fits to the profiles produced by the imaging code.}
  \label{Fig_magfit2010}
\end{figure}

\begin{figure*}
  \centering
  \includegraphics[angle=90,width=\columnwidth]{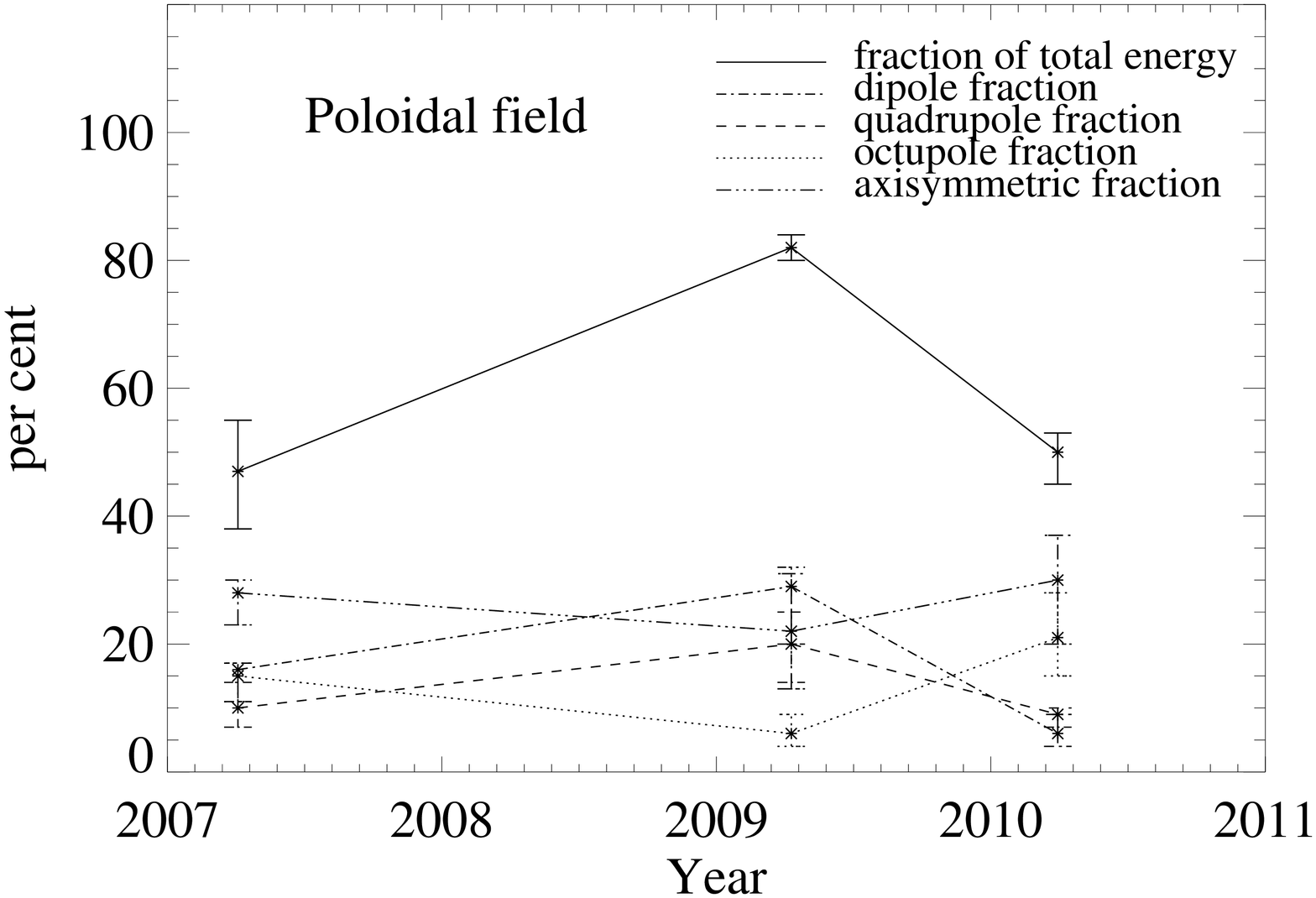}
  \includegraphics[angle=90,width=\columnwidth]{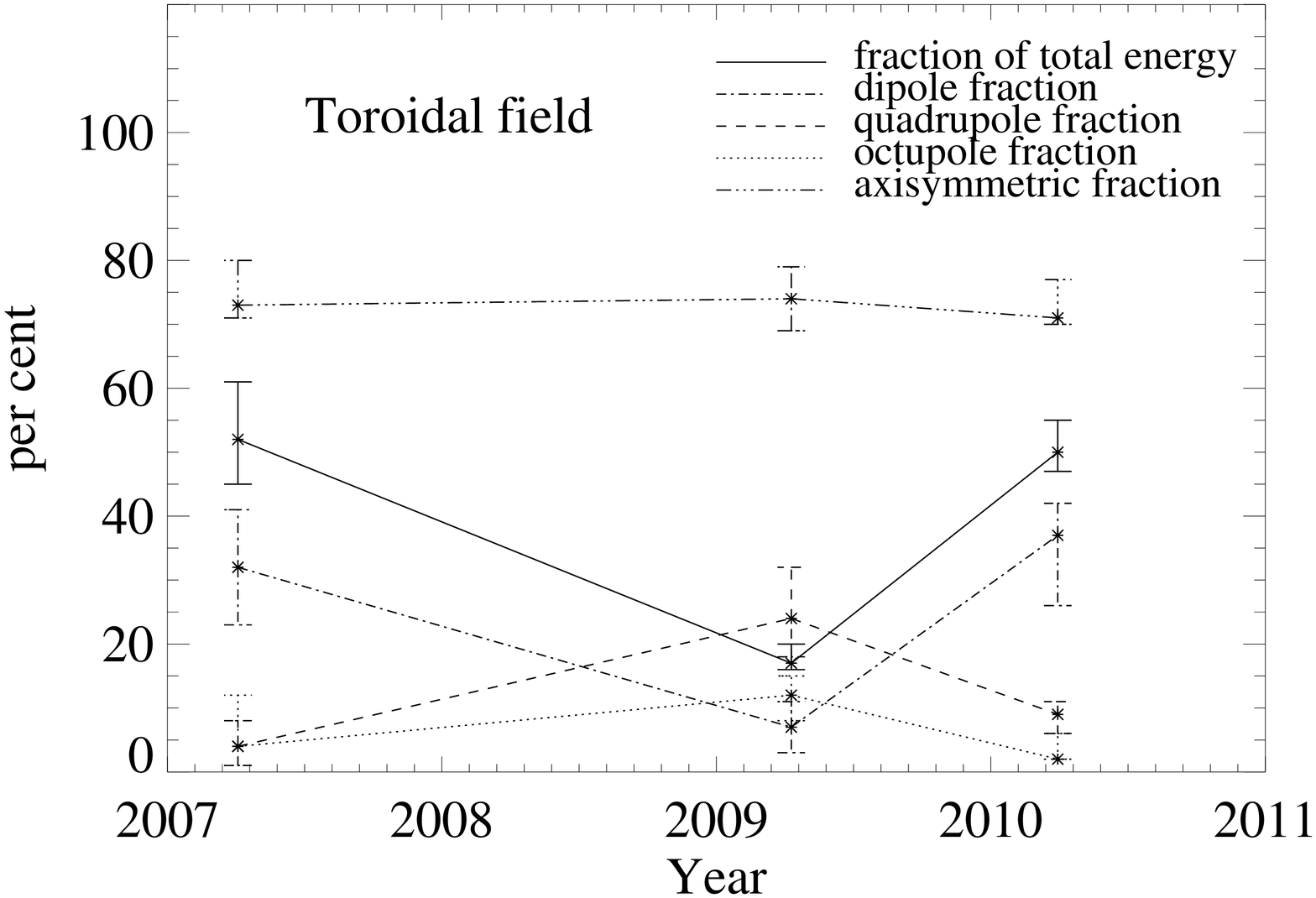}
  \caption{Plot of the magnetic parameters of HD 141943 given in Table~\ref{Tab_magcomp} for the poloidal field (right-hand plot) and toroidal field (left-hand plot). The solid line is the per cent of the large scale energy in the reconstructed poloidal or toroidal field components, the percent of the poloidal or toroidal magnetic energy in dipole (dot-dashed line), quadrupole (dashed line) and octupole (dotted line) components are also given. The percent of the poloidal or toroidal field that is axisymmetric is plotted as a triple-dot-dashed line.}
  \label{Fig_magcomp}
\end{figure*}

The reconstructed magnetic topologies from the March/April 2007, April 2009 and March/April 2010 epochs are given in Fig.~\ref{Fig_allmap2007}, Fig.~\ref{Fig_allmap2009} and Fig.~\ref{Fig_allmap2010} respectively. The fits to the observed Stokes V LSD profiles are given in Fig.~\ref{Fig_magfit2007} (2007 observations), Fig.~\ref{Fig_magfit2009} (2009 observations) and Fig.~\ref{Fig_magfit2010} (2010 observations). All three epochs were fitted to a reduced $\chi^{2}$ value of 1.0. Table~\ref{Tab_magcomp} list the magnetic quantities derived from the magnetic maps in Fig.~\ref{Fig_allmap2007}, Fig.~\ref{Fig_allmap2009} and Fig.~\ref{Fig_allmap2010} and Fig.~\ref{Fig_magcomp} plots these values out. As done in \citet{PetitP:2008} we have varied the input parameters (namely inclination angle, \vsini, rotational period and differential rotation) by the errors given in Table~\ref{Tab_param} and Paper II to determine the possible errors in the magnetic quantities given in Table~\ref{Tab_magcomp}. These errors apply only for the simplified weighting scheme we have used in our reconstructions (see above) and assuming a potential plus toroidal field in the magnetic field reconstruction. They are thus likely to be lower-limits.

\begin{table*}
\caption{Magnetic quantities derived from the magnetic maps in Fig.~\ref{Fig_allmap2007}, Fig.~\ref{Fig_allmap2009} and Fig.~\ref{Fig_allmap2010}. The first column is the epoch of the observation, while the second column is the mean field modulus over the stellar surface. The third and fourth columns are the per cent of the large scale magnetic energy in the reconstructed poloidal and toroidal field components. The fifth, sixth and seventh columns are the per cent of the poloidal magnetic energy in dipole ($l$ $=$ 1), quadrupole ($l$ $=$ 2) and octupole components ($l$ $=$ 3), while the eighth, ninth and tenth columns are the same for the toroidal field. Columns 11 and 12 are the per cent of the poloidal and toroidal field that are axisymmetric ($m$ $=$ 0).}
\label{Tab_magcomp}
\centering 
\begin{tabular}{lccccccccccc}
\hline\hline
                   &                                       & \%                       & \%                          & \% pol.                & \%  pol.              & \% pol.                 & \% tor.                 & \% tor.                 & \% tor.                  & \% axis               & \% axis             \\
Epoch       & $<$B\subs{mod}$>$ & pol. en.               & tor. en.                  & dip.                      & quad.                 & oct.                       & dip.                      & quad.                 & oct.                       & sym. pol.            & sym. tor.            \\
\hline
2007.257 & 91$^{+26}_{-11}$ G  & 47$^{+8}_{-9}$ & 52$^{+9}_{-7}$ & 16$^{+1}_{-5}$ & 10$^{+4}_{-3}$ & 15$^{+2}_{-4}$ & 32$^{+9}_{-9}$ & 4$^{+4}_{-3}$    & 4$^{+8}_{-3}$   & 28$^{+2}_{-5}$ & 73$^{+7}_{-2}$\\
2009.273 & 37$^{+4}_{-4}$ G       & 82$^{+2}_{-2}$ & 17$^{+3}_{-1}$   & 29$^{+2}_{-9}$ & 20$^{+5}_{-6}$ & 6$^{+3}_{-2}$    & 7$^{+4}_{-4}$   & 24$^{+8}_{-6}$ & 12$^{+3}_{-4}$ & 22$^{+10}_{-9}$ & 74$^{+5}_{-5}$\\
2010.244 & 71$^{+19}_{-10}$ G  & 50$^{+3}_{-5}$ & 50$^{+5}_{-3}$   & 6$^{+3}_{-2}$   & 9$^{+1}_{-2}$    & 21$^{+7}_{-6}$ & 37$^{+5}_{-11}$ & 9$^{+2}_{-3}$   & 2$^{+4}_{-0}$    & 30$^{+7}_{-10}$ & 71$^{+6}_{-1}$\\ 
\hline
\end{tabular}
\end{table*}

The reconstructed magnetic topology maps shown in Fig.~\ref{Fig_allmap2007}, Fig.~\ref{Fig_allmap2009} and Fig.~\ref{Fig_allmap2010} show the maximum entropy reconstruction of the large-scale magnetic topology of HD 141943 at the three epochs, 2007.257, 2009.273 and 2010.244. The magnetic imaging code mainly reconstructs the large-scale magnetic topology on the stellar surface of the star, as magnetic flux that is contained in small dipoles below the resolution limit of the observations ($\sim$11\sups{\circ} in longitude at the stellar equator) is likely to cancel out and not be recovered.

As has been previously described \citep[i.e.][]{DonatiJF:1997b} the technique of ZDI sometimes suffers cross-talk between radial and meridional components of the magnetic field for low-latitude features. However, this effect is mainly present in stars with low inclination angles. Given the relatively large inclination angle of HD 141943 ($i$ $\sim$ 70\sups{\circ}) this effect should be minimal. Conversely the sensitivity of ZDI to low-latitude meridional fields decreases significantly with an increase in stellar inclination, and thus the meridional field topologies of HD 141943 (lower-right images of Fig.~\ref{Fig_allmap2007}, Fig.~\ref{Fig_allmap2009} and Fig.~\ref{Fig_allmap2010}) may well miss some meridional field that has not been reconstructed. However, given the lack of meridional field seen on most active stars we believe that if there is any missing field it should be minimal.

In the April 2009 observations, there is a large gap in the observations between  a phase of $\sim$0.45 to $\sim$0.75. This is most likely to have an effect on the recovered radial magnetic field (as ZDI is mostly sensitive to radial field around the phase of the observation and to azimuthal field around 0.2 phase from the observation). This could explain the slightly lower intensity of the radial magnetic features (top-right image of Fig.~\ref{Fig_allmap2009}) at these phases and may well mean that the per cent of poloidal magnetic energy in Table~\ref{Tab_magcomp} is slightly underestimated for the April 2009 observations.

The first thing to notice from Table~\ref{Tab_magcomp} is that the mean magnetic field of the 2007 and 2010 maps is significantly higher than that of the 2009 maps ($\sim$91 G and $\sim$71 G compared to $\sim$37 G). This could either be a significant change in the magnetic field strength on HD 141943 during 2009 or it may well be due to the limited amount (and poorer quality) of observations in the 2009 dataset compared with that of 2007 and 2010. To test this we took only a limited number of profiles from the 2007 and 2010 datasets that had a similar phase coverage to that of the 2009 dataset. This had the result of decreasing the mean magnetic field to $\sim$66 G and $\sim$51 G for the 2007 and 2010 datasets respectively. In addition, we artificially increased the S/N of the Apr 2009 profiles by fitting the data to a reduced $\chi^{2}$ of 0.8, which had the effect of almost doubling the mean magnetic field on the star to $\sim$60 G. Thus we believe that the poor quality of the April 2009 data could well be the reason for the change in mean magnetic field on the star, rather than any actual change in the magnetic field strength of the star.

The other obvious feature of Table~\ref{Tab_magcomp} is the observation that the balance between poloidal and toroidal magnetic energy changed dramatically in 2009 compared to the 2007 and 2010 epochs, being roughly balanced in 2007 and 2010 but being dominated by the poloidal component in 2009. Using the magnetic maps created from a limited number of profiles in 2007 and 2010 (see above) we tested to see if the limited dataset in April 2009 could have affected the ratio of poloidal and toroidal magnetic energy seen in the 2009 observations. Reducing the number of profiles used in the creation of the magnetic map had very little effect on the balance between poloidal and toroidal field in both the 2007 and 2010 datasets (still almost evenly balanced). Thus we believe that the change in the ratio of poloidal and toroidal magnetic energy in the 2009 epoch may well be real, but we cannot rule out the poor dataset in 2009 being responsible.

The major effect that limiting the number of observations in the 2007 and 2010 datasets had was to slightly decrease the complexity of the magnetic field (with more energy in octupole components or lower) and this may explain the slight lowering of the complexity of the poloidal field seen in the 2009 dataset. 

\begin{figure}
  \centering
  \includegraphics[width=\columnwidth]{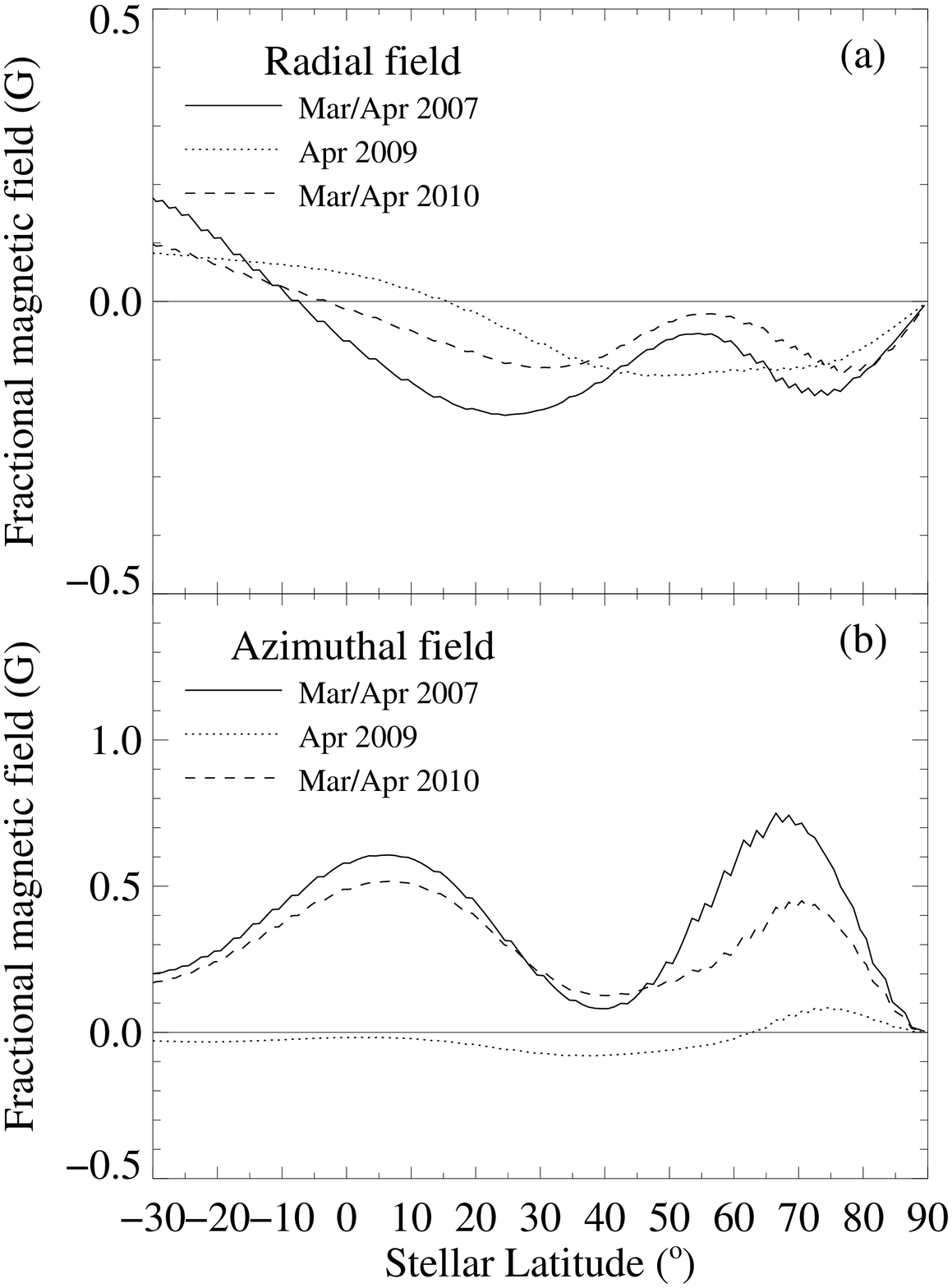}
  \caption{Fractional magnetic field for (a) the radial magnetic field and (b) the azimuthal magnetic field versus stellar latitude for HD141943. This is based on the absolute value of the magnetic field at each latitude and is defined by equation~\ref{Eqn_frac}, with $S(\theta)$ now representing the value of the magnetic field.}
  \label{Fig_frac_mag_lat}
\end{figure}

\begin{figure*}
  \centering
  \includegraphics[angle=90, width=0.86\textwidth]{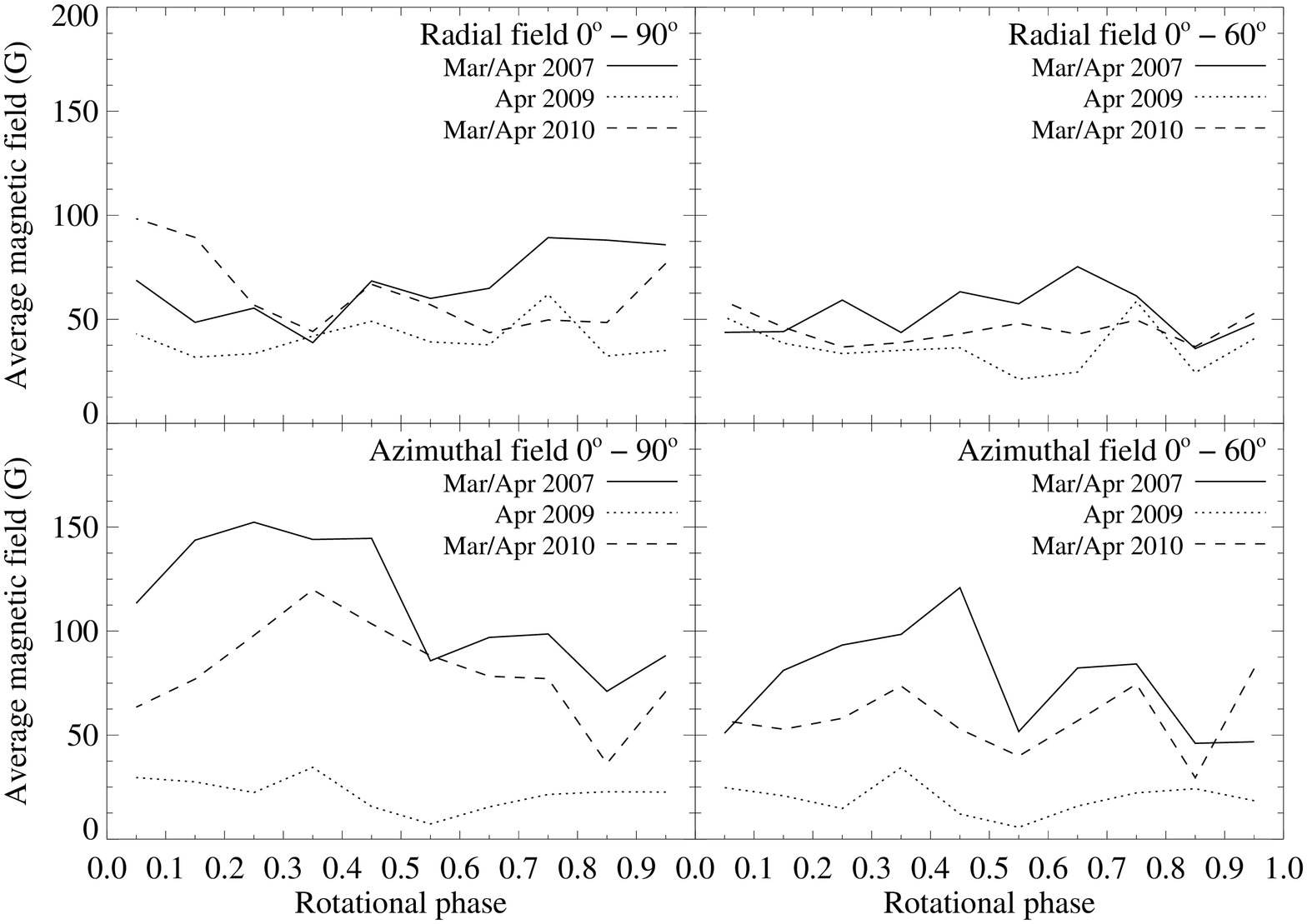}
  \caption{Average of the absolute value of the radial magnetic field (upper plots) and the azimuthal magnetic field (lower plots) versus stellar rotational phase for HD141943, averaged over 0\sups{\circ} to +90\sups{\circ} (left-hand plots) and 0\sups{\circ} to +60\sups{\circ} (right-hand plots) latitude. As was done in Fig.~\ref{Fig_frac_spot_lng}, the phase has been binned into 0.1 steps to remove any small scale variations.}
  \label{Fig_frac_mag_lng}
\end{figure*}

As was done for the spot features, the variation in magnetic field with stellar latitude is given in Fig.~\ref{Fig_frac_mag_lat}. This is again determined using equation~\ref{Eqn_frac}, but with $S(\theta)$ representing the value of the magnetic field at each latitude rather than spot occupancy. We have also determined the variation in magnetic field with longitude, averaging the absolute value of the magnetic field at each longitude. This is shown in Fig.~\ref{Fig_frac_mag_lng}.

Fig.~\ref{Fig_frac_mag_lat} shows that the radial magnetic field of HD 141943  is predominantly negative at all epochs for latitudes above $\sim$0\sups{\circ}. The azimuthal field is predominantly positive at all latitudes in 2007 and 2010 but in 2009 it is positive at high latitudes and predominantly negative at latitudes below $\sim$60\sups{\circ}. Although we feel that this change in the azimuthal field, along with the change in the ratio of poloidal/toroidal field, is possible evidence for a changing magnetic field on HD 141943, we cannot rule out the poor 2009 dataset being responsible. As with the brightness images, Fig.~\ref{Fig_frac_mag_lng} shows little or no evidence of active longitudes in the magnetic fields on HD 141943.

\section{Discussion} \label{Sec_dis}

HD 141943 is only the second \citep[or third including the results for HD 106506 by][]{WaiteIA:2010} young early-G star for which the large-scale magnetic topology has been determined, the other being HD 171488 \citep{MarsdenSC:2006a, JeffersSV:2008, JeffersSV:2010}. However, there have been five young early-G stars for which spot maps and differential rotation measures have been determined (and a number of others that have just spot maps, see Table 4 in \citealt{StrassmeierKG:2003}). Along with HD 141943, HD 171488 and HD 106506, the other two stars are, R58 \citep{MarsdenSC:2005a, MarsdenSC:2005b} and LQ Lup \citep{DonatiJF:2000}. For comparative purposes in this discussion the stellar parameters of all these five stars are given in Table~\ref{Tab_gstars}.

\begin{table*}
\caption{Comparison of the stellar parameters of the five young early-G stars that have had their spot topology and surface differential rotation measured using Doppler imaging. The first three, HD 141943, HD 106506 and HD 171488, have also had their magnetic topologies imaged. Except where noted, the data for HD 141943 are taken from this paper and Paper II, the data for HD 106506 are from \citet{WaiteIA:2010} and that for HD 171488 are from \citet{StrassmeierKG:2003} and \citet{MarsdenSC:2006a}. The data for R58 comes from \citet{MarsdenSC:2005b} and the data for LQ Lup from \citet{DonatiJF:2000}. The values for the depth of the convective zone are from \citet{SiessL:2000}.}
\label{Tab_gstars}
\centering
\begin{tabular}{lccccc}
\hline\hline
Parameter & HD 141943 & HD 106506 & HD 171488 & R58 & LQ Lup \\
\hline
(B-V)                                                          & 0.65$^{a}$                                      & 0.605                                                & 0.62$^{a}$                                      & 0.61$^{b}$                                        & 0.69$^{c}$            \\
Age (Myrs)                                                & $\sim$17                                         & $\la$10                                            & 30 -- 50                                             & 35 $\pm$ 5                                      & 25 $\pm$ 10         \\
Mass (M\subs{\odot})                              & $\sim$1.3                                        & 1.5 $\pm$ 0.1                                 & 1.20 $\pm$ 0.02                              & 1.15 $\pm$ 0.05                            & 1.16 $\pm$ 0.04  \\
Radius (R\subs{\odot})                           & 1.6 $\pm$ 0.1                                 & 2.15 $\pm$ 0.26                             & 1.15 $\pm$ 0.08                             & 1.18$^{+0.17}_{-0.10}$                & 1.22 $\pm$ 0.12   \\
Inclination (\sups{\circ})                          & 70 $\pm$ 10                                   & 65 $\pm$ 5                                      & 60 $\pm$ 10                                   & 60 $\pm$ 10                                    & 35 $\pm$ 5            \\
Convective zone (R\subs{\star})           & $\sim$0.16 [0.26 R\subs{\odot}] & $\sim$0.22 [0.47 R\subs{\odot}] & $\sim$0.21 [0.24 R\subs{\odot}] & $\sim$0.21 [0.25 R\subs{\odot}] & $\sim$0.26 [0.32 R\subs{\odot}] \\
Equatorial rotation period (d)                & $\sim$2.2                                        & $\sim$1.4                                         & $\sim$1.3                                        & $\sim$0.56                                      & $\sim$0.31            \\
\hline
\end{tabular}
\\
$^a$from \citet{CutispotoG:2002}; $^b$from \citet{RandichS:2001}, dereddened value; $^c$from \citet{WichmannR:1997}.
\end{table*}

\subsection{Spot topology} \label{Sec_spot}

The surface spot topology of HD 141943 at four epochs is shown in Fig.~\ref{Fig_map2006}, Fig.~\ref{Fig_allmap2007} (top-left image), Fig.~\ref{Fig_allmap2009} (top-left image) and Fig.~\ref{Fig_allmap2010} (top-left image). The 4 maps show that the spot topology of HD 141943 was remarkably consistent over the span of the observations ($\sim$4 years). All maps show that HD 141943 has a smallish polar spot with a number of lower-latitude features situated predominantly between the equator and +30\sups{\circ} latitude, with only the 2010 epoch showing some significant spot features between +30\sups{\circ} and +60\sups{\circ} latitude at around phase $\sim$0.90. The total spot coverage for all 4 epochs is also very similar ranging from 2.1 per cent to 3.1 per cent.

Comparing these maps to those of other young early-G stars created using the same imaging code, such as R58 \citep{MarsdenSC:2005b}, LQ Lup \citep{DonatiJF:2000}, HD 171488 \citep{MarsdenSC:2006a, JeffersSV:2008, JeffersSV:2010} and HD 106506 \citep{WaiteIA:2010}, it is quite noticeable that while all stars appear to have some lower-latitude features, the polar spot on HD 141943 is significantly smaller than that shown by these other targets. \citet{MarsdenSC:2005b} has shown that starspot mapping assuming high stellar inclination angles leads to a dramatic increase in the amount of polar spot features needed to match the observed deviations in the LSD profiles. Given that HD 141943 has a higher stellar inclination than the other stars listed in Table~\ref{Tab_gstars} it is not an incorrectly determined stellar inclination that is responsible for the small polar spot on HD 141943.

As can be seen in Table~\ref{Tab_gstars}, HD 141943 is the second youngest and second most massive (behind HD 106506) of the young early-G stars so far imaged using this code. It also has the shallowest convective zone (as a function of the stellar radius) and is the slowest rotator of the five. Which of these factors is responsible for the small polar spot seen on HD 141943 is still open for debate. The young early-K star LQ Hya has a rotation rate somewhere between that of HD 141943 and HD 171488 (P $\sim$ 1.6 d) and has a convective zone depth of $\sim$0.29 R\subs{\star}. In some spot maps LQ Hya also show a smallish polar spot \citep{DonatiJF:2003}, although the maps that show this are often those that do not have well sampled observational phases \citep[i.e. Fig. 13,][]{DonatiJF:2003}. In addition, preliminary observations of the late-F star HR 1817 \citep{MarsdenSC:2006b}, with a rotation period of $\sim$1.0 d and a convective zone depth of $\sim$0.17 R\subs{\star}, have shown it to have a small, but intense, polar spot. A significantly larger sample size is required before any firm conclusions can be made about the relationship between polar spot size/intensity and basic stellar parameters. Indeed, given the possible variation in the size of the polar spot seen on LQ Hya \citep{DonatiJF:2003} it may well be that young active stars go through cycles where the size of the polar spot varies as has been mentioned by \citet{BarnesJR:2005}.

\citet{GranzerT:2004} has used thin flux tube models to determine the latitude distribution of spot emergence for young rapidly-rotating stars. Although these models are based on a solar-like interface dynamo (which may not be applicable to such stars, see Section~\ref{Sec_mtop}) it is still interesting to see how well the models compare to the results for HD 141943. Fig.~\ref{Fig_frac_spot_lat} shows that at all 4 epochs HD 141943 has a peak in fractional spottedness around 0\sups{\circ} to +40\sups{\circ} latitude, with another peak around +70\sups{\circ} to +85\sups{\circ}. For a PMS star of 1.0 M\subs{\odot} the models of \citet{GranzerT:2004} predict that the star should have a spot emergence distribution of between +30\sups{\circ} to +70\sups{\circ} latitude with a peak around +50\sups{\circ}. For a 1.7 M\subs{\odot} PMS star this distribution is more spread out with a range from +20\sups{\circ} to +70\sups{\circ} with a peak around +30\sups{\circ} and +40\sups{\circ}. The distribution of fractional spottedness shown in Fig.~\ref{Fig_frac_spot_lat} is not that dissimilar to the combination of the 1.0 M\subs{\odot} and 1.7 M\subs{\odot} distributions given by \citet{GranzerT:2004} with the exception that the higher-latitude peak in spottedness is at a higher latitude in our images than predicted by the models. However, the models predict that for the more Zero-Age Main-Sequence (ZAMS) age stars, such as most of the other stars in Table~\ref{Tab_gstars}, the latitude of peak distribution in spots is predicted to decrease, but the more intense polar spots of these stars would appear to indicate that this is not happening. 

As suggested by \citet{GranzerT:2004} and others, it is possible a strong meridional flow is pushing most of the spots, which form at lower-latitudes, to the polar region, with stars with deeper convective zones possibly having stronger meridional flows and thus forming more intense polar spots. Such a mechanism might well explain the bimodal spot distribution seen in Fig.~\ref{Fig_frac_spot_lat} with the lower peak in the distribution representing the latitude at which the flux tubes emerge and the polar spot those spots that have been driven to the poles by the meridional flow. Such a mechanism may also explain the non-uniform polar spots seen on a number of stars, if the polar spot is being formed by a number of smaller spot features. This is still just speculative, but \citet*{WeberM:2005} have shown tentative evidence for large poleward meridional flows on early-K giants.

\subsection{Large-scale magnetic topology} \label{Sec_mtop}

The maps of the large-scale magnetic topology on HD 141943 from three epochs (Fig.~\ref{Fig_allmap2007}, Fig.~\ref{Fig_allmap2009} and Fig.~\ref{Fig_allmap2010}) show that the radial field on HD 141943 had a mixed polarity at all latitudes for all epochs. The azimuthal field appears to be dominated by a ring of positive field around the pole at all epochs, although the intensity of the polar ring is reduced in 2009. This is shown in the results from Table~\ref{Tab_magcomp}, which show that the toroidal field is predominately axisymmetric ($\sim$70 -- 75 per cent) while the poloidal field is predominately non-axisymmetric ($\sim$70 -- 80 per cent), with the ratio not appearing to change significantly between the three epochs. The other obvious feature of the azimuthal field is the increase in the amount of negative field on the stellar surface in the 2009 image compared to the 2007 and 2010 images. As these surface regions of radial and azimuthal field are believed to be the poloidal and toroidal components of the large-scale dynamo field, the latitudinal distribution and evolution of these regions are an important window onto the operation of the stellar magnetic dynamo.

There appears to be no large-scale magnetic polarity reversal on HD 141943 over the 3 years of observations. This is similar to the young early-G star HD 171488 \citep{MarsdenSC:2006a, JeffersSV:2008, JeffersSV:2010} and the young early-K stars AB Dor and LQ Hya \citep{DonatiJF:2003} none of which have shown evidence of magnetic polarity reversals over several years of observations. In contrast, a magnetic polarity reversal has been seen on the mature Sun-like star HD 190771 \citep{PetitP:2009} in 3 years of observations and the mature late-F star Tau Boo has shown two magnetic polarity reversals in the space of $\sim$2 years \citep{DonatiJF:2008, FaresR:2009}. However, Tau Boo is host to a ``Hot Jupiter'' \citep{ButlerRP:1997} that may be affecting the star's magnetic cycle. 

An axisymmetric field in the form of a ring of azimuthal field around the pole appears to be a common element of active solar-type stars with such a ring also seen on all reconstructed images of the young early-G stars HD 171488 \citep{MarsdenSC:2006a, JeffersSV:2008, JeffersSV:2010} and HD 106506 \citep{WaiteIA:2010}, as well as being seen on a number of images of the young early-K star AB Dor \citep{DonatiJF:2003}. Such a ring has also been seen on more evolved solar-type stars such as the early-G FK Com star HD 199178 \citep{PetitP:2004b} and the early-K RSCVn star HR 1099 \citep{DonatiJF:2003, PetitP:2004a}. Such large-scale regions of azimuthal magnetic field near the surface of these active stars has lead to the belief that such stars have a fundamentally different dynamo in operation than that of the solar-type interface-layer dynamo, as such regions should not be seen near the stellar surface if a solar dynamo were in action. It is thought that such stars contain a distributed dynamo, one that operates throughout the stellar convective zone rather than being restricted to the interface layer \citep[see][]{DonatiJF:2003}.

Such a concept has recently been supported by theoretical models of \citet{BrownBP:2010} which have shown that a stable magnetic dynamo can exist without an interface-layer. These models are based on a rapidly-rotating Sun (up to 3 times the current solar rotation rate) and also show large regions of near-surface ``wreaths'' of longitudinal magnetic field. These wreaths are similar to the rings of azimuthal field seen on active solar-type stars, except that they are at lower latitudes. Perhaps the higher Coriolis force experienced by the more rapidly rotating stars such as HD 141943 (rotating around 10 times the solar value) are pushing these wreaths to higher latitudes.

Table~\ref{Tab_magcomp} shows that in all epochs both the poloidal and toroidal magnetic fields are very complex with over 50 per cent of the magnetic energy in orders higher than an octupole (except for the poloidal magnetic energy in 2009 which has only 45 per cent of the magnetic energy in orders higher than an octupole). There also appears to be evolution in the magnetic field between the epochs. In 2007 the magnetic energy is reasonably evenly balanced between poloidal and toroidal components (47 per cent to 52 per cent respectively) while in 2009 the poloidal field is extremely dominant (82 per cent to 17 per cent) and in 2010 the magnetic energy is again balanced between the poloidal and toroidal components (50 per cent in each). As explained in Section~\ref{Sec_mag}, the dataset in 2009 is significantly poorer than those in 2007 and 2010 and this may be having an influence on the reconstruction of the magnetic topology for this epoch. However, when we tried to reconstruct the magnetic topology for the 2007 and 2010 epochs using a small subset of the data similar to that of the 2009 dataset there was virtually no change in the ratio of poloidal and toroidal field. Thus we feel that the change in the poloidal/toroidal ratio seen in 2009 may well be real, but unfortunately we cannot prove this with any certainty.

\citet{DonatiJF:2009} has looked at the changing magnetic field topologies on stars of different masses and rotation rates \citep[see Fig. 3 in][]{DonatiJF:2009}. This shows that stars around the mass and rotation rate of HD 141943 (actually preliminary results from the March/April 2007 dataset on HD 141943 are plotted in the figure) fall into an `active' dynamo regime, which includes stars with Rossby number $\la$ 1 and more massive than $\sim$0.5 M\subs{\odot}. These stars produce predominantly non-axisymmetric poloidal magnetic field topologies as well as having significant amounts of toroidal field. At all three epochs studied in this paper HD 141943 does appear to have a predominantly non-axisymmetric poloidal field (and a predominantly axisymmetric toroidal field), but, as mentioned above, the ratio of poloidal to toroidal magnetic field appears to change for the 2009 epoch.

A change in the ratio of poloidal to toroidal magnetic field has been seen on another star. The previously mentioned mature late-F star Tau Boo (period $\sim$3 days) shows large changes in its ratio of poloidal/toroidal field during a magnetic cycle, with toroidal field dominating before a global polarity switch and poloidal field dominating after \citep{FaresR:2009}. This could indicate that the changes seen in the ratio of poloidal and toroidal field on HD 141943 are part of a stellar magnetic cycle on the star, however it would mean that it has undergone an extremely rapid evolution of its magnetic field, but has yet to undergo a magnetic polarity reversal. It would be useful to follow HD 141943 on at least yearly timescales to see whether the magnetic field does in fact undergo rapid magnetic evolution and polarity reversals, such as that evidenced by Tau Boo.

As part of the Sun's magnetic cycle, the ratio of dipole to quadrupole components of the magnetic field changes. As can been seen in Table~\ref{Tab_magcomp} there are significant changes in the percent of magnetic energy in dipole, quadrupole and octupole components for both the poloidal and toroidal fields. For example the per cent of poloidal magnetic energy in the dipole component changes from $\sim$6 per cent in 2007 to $\sim$29 per cent in 2009 and down to $\sim$6 per cent in 2010. The most significant changes appear to occur for the 2009 dataset. Assuming that the change seen in the magnetic topology of HD 141943 in 2009 is real, a comparison of the total magnetic energy (both poloidal and toroidal components) of HD 141943 appears to show that HD 141943 has also undergone a change in the complexity of its magnetic field from 2007 to 2009 and then back again in 2010. From Table~\ref{Tab_magcomp} the total dipole component of the magnetic energy stays reasonably constant over all three epochs ($\sim$24 per cent in 2007, $\sim$25 per cent in 2009 and $\sim$22 per cent in 2010). The same can be said for the total octupole component ($\sim$9 per cent in 2007, $\sim$7 per cent in 2009 and $\sim$12 per cent in 2010). However, the total quadrupole component appears to change markedly in 2009 compared to the other two epochs ($\sim$7 per cent in 2007, $\sim$18 per cent in 2009 and $\sim$9 per cent in 2010) mostly due to an increase in the quadrupole component of the poloidal field. Now, as explained in Section~\ref{Sec_mag} the poorer observations in 2009 may play a role in this change but this could be a change imparted by magnetic evolution on HD 141943.

\section{Conclusions} \label{Sec_con}

HD 141943 is one of the youngest and most massive stars for which we have analysed the surface magnetic topologies for. For young stars it has the shallowest convective zone that we have yet studied. We have presented reconstructed brightness (at 4 epochs) and magnetic (at 3 epochs) topologies of HD 141943. During the four year timebase of the brightness images the spot distribution on HD 141943 has changed very little with a smallish polar spot and a number of lower-latitude features at around 0\sups{\circ} to +30\sups{\circ} latitude, seen at all epochs. 

Over three years of observations the pattern of the magnetic topology of HD 141943 also looks at first glance to have experienced relatively little change, with positive and negative radial magnetic field seen at all latitudes and a ring of positive azimuthal magnetic field seen at all epochs. At all epochs the large-scale poloidal magnetic field of HD 141943 is mostly non-axisymmetric while the toroidal field is predominantly axisymmetric. The reconstructed magnetic topologies are rather complex with over 50 per cent of the magnetic energy in components higher than an octupole. When the magnetic images were analysed in more detail we found tentative evidence for a change in the magnetic field in 2009. The ratio of poloidal to toroidal field on HD 141943 goes from almost balanced in 2007 to being heavily dominated by poloidal magnetic field in 2009 and back to balanced in 2010. If real, this variation would indicate magnetic evolution on HD 141943.

\section*{Acknowledgments}

The observations in this paper were obtained with the Anglo-Australian telescope. We would like to thank the technical staff of the Anglo-Australian Observatory (now the Australian Astronomical Observatory) for their, as usual, excellent assistance during these observations. We would also like to thank the anonymous referee who helped improve this paper. This project is supported by the Commonwealth of Australia under the International Science Linkages program.



\label{lastpage}


\begin{thebibliography}{}
\bibitem[\protect\citeauthoryear{Baliunas et al.}{1995}]{BaliunasSL:1995} Baliunas S. L., Donahue R. A., Soon W. H., et al., 1995, ApJ, 438, 269
\bibitem[\protect\citeauthoryear{Barnes et al.}{2000}]{BarnesJR:2000} Barnes J. R., Collier Cameron A., James D. J., Donati J.-F., 2000, MNRAS, 314, 162
\bibitem[\protect\citeauthoryear{Barnes et al.}{2001a}]{BarnesJR:2001a} Barnes J. R., Collier Cameron A., James D. J., Donati J.-F., 2001a, MNRAS, 324, 231
\bibitem[\protect\citeauthoryear{Barnes et al.}{2001b}]{BarnesJR:2001b} Barnes J. R., Collier Cameron A., James D. J., Steeghs D.., 2001b, MNRAS, 326, 1057
\bibitem[\protect\citeauthoryear{Barnes}{2005}]{BarnesJR:2005} Barnes J. R., 2005, MNRAS, 364, 137
\bibitem[\protect\citeauthoryear{Berdyugina}{2005}]{BerdyuginaSV:2005} Berdyugina S. V., 2005, Starspots: A Key to the Stellar Dynamo, Living Rev. Solar Phys. 2, (2005), 8. URL (cited on 2010 January 22): http://www.livingreviews.org/lrsp-2005-8
\bibitem[\protect\citeauthoryear{Berdyugina \& Tuominen}{1998}]{BerdyuginaSV:1998} Berdyugina S. V., Tuominen I., 1998, A\&A, 336, L25
\bibitem[\protect\citeauthoryear{Brandenburg}{2005}]{BrandenburgA:2005} Brandenburg A., 2005, ApJ, 625, 539
\bibitem[\protect\citeauthoryear{Brown et al.}{2010}]{BrownBP:2010} Brown B. P., Browning M. K., Brun A. S., Miesch M. S., Toorme J., 2010, ApJ, 711, 424
\bibitem[\protect\citeauthoryear{Brown et al.}{1991}]{BrownSF:1991} Brown S. F., Donati J.-F., Rees D. E., Semel M., 1991, A\&A, 250, 463
\bibitem[\protect\citeauthoryear{Browning}{2008}]{BrowningMK:2008} Browning M. K., 2008, ApJ, 676, 1262
\bibitem[\protect\citeauthoryear{Butler et al.}{1997}]{ButlerRP:1997} Butler R. P., Marcy G. W., Williams E., Hauser H., Shirts P., 1997, ApJ, 474, L115
\bibitem[\protect\citeauthoryear{Charbonneau}{2005}]{CharbonneauP:2005} Charbonneau P., 2005, Dynamo Models of the Solar Cycle, Living Rev. Solar Phys. 2,  (2005),  2. URL (cited on 2010 August 19): http://www.livingreviews.org/lrsp-2005-2
\bibitem[\protect\citeauthoryear{Collier Cameron}{1992}]{CameronAC:1992} Collier Cameron A., 1992, in Byrne P. B., Mullan D. J., eds, Lecture Notes in Physics, Vol. 397, Surface Inhomogeneities on Late-Type Stars, Springer, Berlin, p. 33 
\bibitem[\protect\citeauthoryear{Cutispoto et al.}{1999}]{CutispotoG:1999} Cutispoto G., Pastori L., Tagliaferri G., Messina S., Pallavicini R., 1999, A\&AS, 138, 87
\bibitem[\protect\citeauthoryear{Cutispoto et al.}{2002}]{CutispotoG:2002} Cutispoto G., Pastori L., Pasquini L., de Medeiros J. R., Tagliaferri G., Anderson J., 2002, A\&A, 384, 491
\bibitem[\protect\citeauthoryear{Cutispoto et al.}{2003}]{CutispotoG:2003} Cutispoto G., Tagliaferri G., de Medeiros J. R., Pastori L., Pasquini L., Anderson J., 2003, A\&A, 397, 987
\bibitem[\protect\citeauthoryear{Donati et al.}{1992}]{DonatiJF:1992} Donati J.-F., Brown S. F., Semel M., Rees D. E., Dempsey R. C., Matthews J. M., Henry G. W., Hall D. S., 1992, A\&A, 265, 682
\bibitem[\protect\citeauthoryear{Donati \& Collier Cameron}{1997}]{DonatiJF:1997a} Donati J.-F., Collier Cameron A., 1997, MNRAS, 291, 1
\bibitem[\protect\citeauthoryear{Donati \& Brown}{1997}]{DonatiJF:1997b} Donati J.-F., Brown S. F., 1997, A\&A, 326, 1135
\bibitem[\protect\citeauthoryear{Donati et al.}{1997}]{DonatiJF:1997c} Donati J.-F., Semel M., Carter B. D., Rees D. E., Cameron A. C., 1997, MNRAS, 291, 658
\bibitem[\protect\citeauthoryear{Donati}{1999}]{DonatiJF:1999b} Donati J.-F., 1999, MNRAS, 302, 457
\bibitem[\protect\citeauthoryear{Donati et al.}{1999}]{DonatiJF:1999a} Donati J.-F., Collier Cameron A., Hussain G., Semel M., 1999, MNRAS, 302, 437
\bibitem[\protect\citeauthoryear{Donati et al.}{2000}]{DonatiJF:2000} Donati J.-F., Mengel M., Carter B. D., Marsden S., Collier Cameron A., Wichmann R., 2000, MNRAS, 316, 699
\bibitem[\protect\citeauthoryear{Donati et al.}{2003}]{DonatiJF:2003} Donati J.-F., Collier Cameron A., Semel M., et al., 2003, MNRAS, 345, 1145
\bibitem[\protect\citeauthoryear{Donati et al.}{2006}]{DonatiJF:2006} Donati J.-F., Howarth I. D., Jardine M. M., et al., 2006, MNRAS, 370, 629
\bibitem[\protect\citeauthoryear{Donati et al.}{2008}]{DonatiJF:2008} Donati J.-F., Moutou C., Far\`{e}s R., et al. 2008, MNRAS, 385, 1179
\bibitem[\protect\citeauthoryear{Donati \& Landstreet}{2009}]{DonatiJF:2009} Donati J.-F., Landstreet J. D., 2009, ARA\&A, 47, 333
\bibitem[\protect\citeauthoryear{Dunstone et al.}{2008}]{DunstoneNJ:2008} Dunstone H. J., Hussain G. A. J., Collier Cameron A., Marsden S. C., Jardine M., Stempels H. C., Ramirez Vlez J. C., Donati J.-F., 2008, MNRAS, 387, 481
\bibitem[\protect\citeauthoryear{Fares et al.}{2009}]{FaresR:2009} Fares R., Donati J.-F., Moutou C., et al., 2009, MNRAS, 398, 1383
\bibitem[\protect\citeauthoryear{Granzer et al.}{2000}]{GranzerT:2000} Granzer Th. Sch\"{u}ssler M., Caligari P., Strassmeier K. G., 2000, A\&A, 355, 1087
\bibitem[\protect\citeauthoryear{Granzer}{2004}]{GranzerT:2004} Granzer T., 2004, Astron. Nachr., 325, 417
\bibitem[\protect\citeauthoryear{Hillenbrand et al.}{2008}]{HillenbrandLA:2008} Hillenbrand L. A., Carpenter J. M., Kim J. S., et al., 2008, ApJ, 677, 630
\bibitem[\protect\citeauthoryear{J\"{a}rvinen et al.}{2005}]{JarvinenSP:2005} J\"{a}rvinen S. P., Berdyugina S. V., Tuominen I., Cutispoto G., Bos M., 2005, A\&A, 432, 657
\bibitem[\protect\citeauthoryear{Jeffers \& Donati}{2008}]{JeffersSV:2008} Jeffers S. V., Donati J.-F., 2008, MNRAS, 390, 635
\bibitem[\protect\citeauthoryear{Jeffers et al.}{2010}]{JeffersSV:2010} Jeffers S. V., Donati J.-F., Alecian E., Marsden S. C., 2010, MNRAS, accepted
\bibitem[\protect\citeauthoryear{Kochukhov, Makaganiuk \& Piskunov}{Kochukhov}{2010}]{KochukhovO:2010} Kochukhov O., Makaganiuk V., Piskunov N., 2010, A\&A, 524, A5
\bibitem[\protect\citeauthoryear{Kurucz}{1993}]{KuruczRL:1993} Kurucz R. L., 1993, CDROM \#13 (ATLAS9 atmospheric models) and CDROM \#18 (ATLAS9 and SYNTHE routines, spectral line database)
\bibitem[\protect\citeauthoryear{Mackay et al.}{2004}]{MackayDH:2004} Mackay D. H., Jardine M., Collier Cameron A., Donati J.-F., Hussain G. A. J., 2004, MNRAS, 354, 737
\bibitem[\protect\citeauthoryear{Marsden et al.}{2005a}]{MarsdenSC:2005a} Marsden S. C., Carter B. D., Donati J.-F., 2005a, in Favata F., Hussain G. A. J., Battrick B., eds. Proceedings of the 13\sups{th} Cambridge Workshop on Cool Stars, Stellar Systems and the Sun, ESA Special Publications, Vol. 560, ESA, Noordwijk, The Netherlands, p. 799
\bibitem[\protect\citeauthoryear{Marsden et al.}{2005b}]{MarsdenSC:2005b} Marsden S. C., Waite I. A., Carter B. D., Donati J.-F., 2005b, MNRAS, 359, 711
\bibitem[\protect\citeauthoryear{Marsden et al.}{2006a}]{MarsdenSC:2006a} Marsden S. C., Donati J.-F., Semel M., Petit P., Carter B. D., 2006a, MNRAS, 370, 468
\bibitem[\protect\citeauthoryear{Marsden et al.}{2006b}]{MarsdenSC:2006b} Marsden S. C., Mengel M. W., Donati J.-F., Carter B. D., Semel M., Petit P., 2006b, in Casini R., Lites B. W., eds. Proceedings of the 4\sups{th} Solar Polarization workshop, ASP Conference Series, Vol. 358, ASP, San Francisco, USA, p. 401
\bibitem[\protect\citeauthoryear{Marsden et al.}{2010a}]{MarsdenSC:2010a} Marsden S. C., Jeffers S. V., Donati J.-F., Mengel M. W., Waite I. A., Carter B. D., 2010a, in Kosovichev A., Rozelot J.-P., Andrei A., eds., Proceedings of IAU Symposium 264, Solar and Stellar Variability: Impact on Earth and Planets, Cambridge University Press, Cambridge, UK, p. 130
\bibitem[\protect\citeauthoryear{Marsden et al.}{2010b}]{MarsdenSC:2010b} Marsden S. C., Jardine M. M., Ram\'{i}rez V'{e}lez J. C., et al., 2010b, MNRAS, accepted (Paper II)
\bibitem[\protect\citeauthoryear{Mengel}{2006}]{MengelM:2006} Mengel M., 2006, Mphil thesis, University of Southern Queensland, Toowoomba, Queensland, Australia
\bibitem[\protect\citeauthoryear{Nordstr\"{o}m et al.}{2004}]{NordstromB:2004} Nordstr\"{o}m B., Mayor M., Andersen J., et al., 2004, A\&A, 418, 989
\bibitem[\protect\citeauthoryear{Parker et al.}{1993}]{ParkerEN:1993} Parker E. N., 1993, ApJ, 408, 707
\bibitem[\protect\citeauthoryear{Petit et al.}{2004a}]{PetitP:2004a} Petit P., Donati J.-F., Wade G. A., et al., 2004a, MNRAS, 348, 1175
\bibitem[\protect\citeauthoryear{Petit et al.}{2004b}]{PetitP:2004b} Petit P., Donati J.-F., Oliveira J. M., et al., 2004b, MNRAS, 351, 826
\bibitem[\protect\citeauthoryear{Petit et al.}{2008}]{PetitP:2008} Petit P., Dintrans B., Solanki S. K., et al., 2008, MNRAS 388, 80
\bibitem[\protect\citeauthoryear{Petit et al.}{2009}]{PetitP:2009} Petit P., Dintrans B., Morgenthaler A., Van Grootel V., Morin J., Lanoux J., Auri\`{e}re M., Konstantinova-Antova R., 2009, A\&A, 508, L9 
\bibitem[\protect\citeauthoryear{Randich}{2001}]{RandichS:2001} Randich S., 2001, A\&A, 377, 512
\bibitem[\protect\citeauthoryear{Schussler et al.}{1996}]{SchusslerM:1996} Sch\"{u}ssler M., Caligari P., Ferriz-Mas A., Solanki S. K., Stix M., 1996, A\&A, 314, 503
\bibitem[\protect\citeauthoryear{Semel}{1989}]{SemelM:1989} Semel M., 1989, A\&A, 225, 456
\bibitem[\protect\citeauthoryear{Semel, Donati \& Rees}{Semel et al.}{1993}]{SemelM:1993} Semel M., Donati J.-F., Rees D. E., 1993, A\&A, 278, 231
\bibitem[\protect\citeauthoryear{Siess, Dufour \& Forestini}{Siess et al.}{2000}]{SiessL:2000} Siess L., Dufour E., Forestini M., 2000, A\&A, 358, 593
\bibitem[\protect\citeauthoryear{Skilling \&  Bryan}{1984}]{SkillingJ:1984} Skilling J., Bryan R. K., 1984, MNRAS, 211, 111
\bibitem[\protect\citeauthoryear{Strassmeier et al.}{2003}]{StrassmeierKG:2003} Strassmeier K. G., Pichler T., Weber M., Granzer T., 2003, A\&A, 411, 595
\bibitem[\protect\citeauthoryear{Unruh \& Collier Cameron}{1995}]{UnruhYC:1995} Unruh Y. C., Collier Cameron A., 1995, MNRAS, 273, 1
\bibitem[\protect\citeauthoryear{Waite et al.}{2005}]{WaiteIA:2005} Waite I. A., Carter B. D., Marsden S. C., Mengel M. W., 2005, PASA, 22, 29
\bibitem[\protect\citeauthoryear{Waite et al.}{2010}]{WaiteIA:2010} Waite I. A., Marsden S.C., Carter B.D., et al., 2010, MNRAS, accepted
\bibitem[\protect\citeauthoryear{Weber, Strassmeier \& Washuettl}{Weber et al.}{2005}]{WeberM:2005} Weber M., Strassmeier K. G., Washuettl A., 2005, Astron. Nachr., 326, 287
\bibitem[\protect\citeauthoryear{Wichmann et al.}{1997}]{WichmannR:1997} Wichmann R., Krautter J., Covino E., Alcal\'{a} J. M., Neuh\"{a}user R., Schmitt J. H. M. M., 1997, A\&A, 320, 185
\end{thebibliography}
\end{document}